%% file: Main_text.tex
\documentclass[twocolumn]{aastex631}

\usepackage{comment}
\usepackage{float}
\usepackage{placeins}
\usepackage{chngcntr}
\usepackage{natbib}
\usepackage{graphicx}
\usepackage{tabularx}
\usepackage{xcolor}

\usepackage{booktabs}
\usepackage{array}
\usepackage{makecell} 
\usepackage{ragged2e} 
\usepackage{threeparttable}

\usepackage{rotating}
\usepackage[T1]{fontenc}

\newcolumntype{L}[1]{>{\RaggedRight\arraybackslash}m{#1}}
\newcolumntype{C}[1]{>{\centering\arraybackslash}m{#1}}

\newcommand{\teff}[1]{$T_{\rm eff}${#1}}

\defcitealias{Jahandar2024}{J24}

\begin{document}

\title{Elemental Abundances of Cool Stars: A Spectroscopic Framework Applied to Five Planet-Host Stars}

\author[0009-0000-1424-7694]{Nicole Gromek}
\affiliation{Department of Physics \& Astronomy, McMaster University, 1280 Main St West, Hamilton, ON L8S 4L8, Canada}

\author[0000-0002-7992-469X]{Drew Weisserman}
\affiliation{Department of Physics \& Astronomy, McMaster University, 1280 Main St West, Hamilton, ON L8S 4L8, Canada}

\author[0000-0001-5383-9393]{Ryan Cloutier}
\affiliation{Department of Physics \& Astronomy, McMaster University, 1280 Main St West, Hamilton, ON L8S 4L8, Canada}

\author[0009-0005-6135-6769]{Alexandrine L'Heureux}  
\affiliation{Institut Trottier de recherche sur les exoplan\`etes, D\'epartement de Physique, Universit\'e de Montr\'eal, Montr\'eal, QC H2V 0B3, Canada}

\author[0000-0001-9291-5555]{Charles Cadieux}  
\affiliation{Observatoire de Gen\`eve, D\'epartement d'Astronomie, Universit\'e de Gen\`eve, Chemin Pegasi 51, 1290 Versoix, Switzerland}
\affiliation{Institut Trottier de recherche sur les exoplan\`etes, D\'epartement de Physique, Universit\'e de Montr\'eal, Montr\'eal, QC H2V 0B3, Canada}

\begin{abstract}
Stellar elemental abundances provide important constraints on the chemical environments from which planets form. M and late-K dwarfs are prime targets for the characterization of small exoplanets, yet their complex spectra make accurate elemental abundance measurements challenging. In this work, we present a spectral synthesis-based framework to derive detailed abundances for cool stars from high-resolution, near-infrared spectra. We apply our framework to SPIRou $YJHK$ spectra of five planet-host stars--GJ~9827, K2-18, TOI-1201, TOI-1685, and Barnard's Star--and derive abundances of C, O, Na, Mg, Al, Si, K, Ca, Ti, Cr, Mn, and Fe for all five stars, plus Sc, V, and Ni for a subset, with typical uncertainties of $\lesssim 0.1$~dex. Our recovered metallicities and elemental abundances are broadly consistent with expected trends and previous literature analyses of these stars. We derive stellar ratios of C/O, Mg/Si, Fe/Mg, and Fe/O and find that all five systems are broadly consistent with solar ratios, with the exception of the metal-poor Barnard's Star ([M/H]$=-0.33\pm0.06$~dex) that is substantially iron-poor ([Fe/H]$=-0.722\pm 0.085$~dex). Our results establish stellar chemical baselines for interior structure and atmospheric studies of these planetary systems with the Hubble Space Telescope (HST) and James Webb Space Telescope (JWST), while highlighting the remaining model-dependent limitations of precise abundance analyses for cool stars.
\end{abstract}


\section{Introduction} \label{sec:intro}
Planets form from the same primordial nebular material as their host stars, establishing a fundamental chemical link between stellar and planetary composition. Consequently, stellar elemental abundances are commonly used as proxies for the initial chemical inventory available for planet formation in the protoplanetary disk \citep{Bond_2010, Dorn_2015, Thiabaud2015, Dorn_2017, Adibek_2021, Brinkman2024a}. A clear demonstration of this planet-star connection is the well-established correlation between stellar metallicity and giant planet occurrence around Sun-like stars \citep{Gonzalez_1997, FischerValenti_2005, Guillot_2006, Johnson_2010, Osborn_2020} and M dwarfs \citep{Gan_2025}, but efforts to find an analogous relation between stellar metallicity and planetary properties for low-mass planets reveal weaker connections \citep[e.g.][]{Bitsch2020, Adibek_2021, Kutra2021, Brinkman2024b, Wanderley_2025, Weisserman_2026, Plonykov2026, Turtelboom_2026}. This suggests that bulk metallicity alone is insufficient to capture the full complexity of small planet formation outcomes. These results have motivated ongoing efforts to probe the role of individual elemental abundances in planet formation \citep{Gilli_2006}. Refractory elements such as Mg, Si, and Fe are fundamental building blocks of rocky planets, dominating super-Earth bulk compositions by way of their primarily silicate mantles and predominantly iron cores \citep{MorganAnders_1980, Adibek_2021}. If stellar and planetary compositions are expected to be similar, then stellar Fe/Mg, Mg/Si, and Fe/O ratios can directly inform the core-to-mantle ratios and internal structures of their planets, offering constraints that help mitigate degeneracies in bulk composition inferences from mass, radius, and temperature measurements alone \citep{RogersSeager_2010, BrewerFischerMgSiCO2016, UnderbornPanero2017, Dorn_2017, Lichtenberg2021}. Volatile element ratios, such as C/O and N/O, provide insight into gas-rich planets' formation locations and migration histories via the relative accretion of gas and solids within the disk \citep{Oberg2011, Madhu2019, Cridland_2020}. In old, iron-poor systems, the relative enhancement of alpha elements may facilitate planet formation by compensating for a reduced iron inventory, thereby dominating the planetary composition \citep{Adibek2012b, Kobayashi2020, Ghezzi_2026}.

M dwarfs are particularly important targets for studying the star-planet compositional connection and tracing Galactic chemical evolution. They represent the most common stars in the Galaxy and their long, stable main-sequence lifetimes make them valuable tracers of chemical enrichment. They host small planets at higher rates than FGK stars \citep[e.g.][]{Henry_2018, DressingChar2015, Mulders2018, Gillis2026}, while their small radii and low masses allow for larger transit and radial velocity signatures from orbiting planets. These properties make M dwarf planetary systems some of the most accessible laboratories for precise characterizations of super-Earths and sub-Neptunes, positioning them as attractive targets for exoplanet discovery surveys and detailed atmospheric characterization efforts \citep{Lustig-Yaeger_2023}. Unfortunately, while chemical trends among FGK planet hosts have been extensively investigated 
\citep[e.g.][]{Delgado2010, Jofre_2015, BrewerFischerMgSiCO2016, DelgadoMena2021, Kolecki_2022, Polanski_2022, Adibek_2024, Brinkman2024b}, comparable population-level chemical studies of planet-hosting M dwarfs remain limited. The principal difficulty lies in the complex spectra of M dwarfs. Their cool atmospheres produce spectra dominated by strong molecular bands from TiO, VO, and H$_2$O \citep{Allard2000}. These features obscure the true continuum and blend with atomic features, thereby limiting the utility of conventional equivalent-width-based methods used on FGK stars to recover elemental abundances. While observing M dwarfs in near-infrared (NIR) wavelengths offers some key advantages over the optical, such as higher flux and less  molecular absorption, incomplete atomic and molecular linelists, telluric absorption, stellar activity, and uncertainties in atmospheric models all continue to impede abundance accuracy \citep{Onehag_2012, Jahandar2024}. As such, obtaining precise and accurate elemental abundances from M dwarf spectra remains an unsolved problem in the field.

Recent advances in high-resolution NIR spectroscopy have begun to make detailed M dwarf elemental abundance analyses possible. Early work used APOGEE $H$-band spectra ($R\sim 22,500$) to derive abundances for a broad range of elements, including C, O, Na, Mg, Al, Si, K, Ca, Ti, V, Cr, Mn, and Fe \citep{Souto_2017}, with subsequent studies extending these methods to larger and more diverse samples \citep{Sarmento2021,Souto_2022,Melo2024,Behmard_2025}. However, the high density of atomic and molecular features in cool star spectra leads to extensive blending, making high spectral resolution critical for isolating individual line profiles. As a result, recent analyses have increasingly relied on high-resolution NIR spectrographs with $R\gtrsim 40,000$, such as Subaru/IRD \citep[e.g.][]{Ishikawa_2020, Ishikawa2022}, CARMENES \citep[e.g.][]{Shan_2021}, IGRINS \citep[e.g.][]{Hejazi2023,Hejazi_2024} and SPIRou \citep{Jahandar2024,Jahandar_2025}, together with spectral synthesis methods to measure cool star elemental abundances. 
The combination of high-resolution NIR spectra, improved telluric-correction techniques, and advances in atmospheric modelling for cool stars has enabled abundance determinations with typical levels of precision of 0.1–0.2~dex.

In this work, we describe our dedicated abundance analysis framework optimized for high-resolution NIR spectra of cool stars. Our analysis builds on recent progress in M dwarf abundance studies \citep[e.g.][]{Souto_2022,Hejazi_2024} and incorporates customized procedures to address local pseudo-continuum normalization, line selection, feature blending, and uncertainty estimation. Our framework has already been applied to several cool planet-host stars in atmospheric and interior characterization studies using data from 
NIRPS \citep{Osborn_2026,Srivastava_2026,Weisserman_2026}, CARMENES-NIR channel \citep{Turner_2026}, and SPIRou (Piaulet-Ghorayeb et al., in prep). Here, we apply it to high-resolution SPIRou spectra of five cool planet-hosting stars: GJ~9827, K2-18, TOI-1201, TOI-1685, and Barnard's Star
, whose planetary systems span a range of architectures including rocky ultra-short-period planets, volatile-rich sub-Neptunes, and terrestrial sub-Earths. 

Our paper is organized as follows. In Section~\ref{sec:stars}, we outline our stellar sample and highlight the scientific context of their planetary systems.  Section~\ref{sec:spirou} describes the SPIRou instrument and our observing program. In Section~\ref{sec:methods}, we detail our abundance analysis methodology. Section~\ref{sec:results} presents our derived elemental abundances and ratios, while Section~\ref{sec:discussion} presents a discussion of our results, including implications for planet formation and characterization and comparisons to literature results.  Section~\ref{sec:summary} concludes with a summary of our main results.

\section{Stellar Sample} \label{sec:stars}
Our stellar sample consists of five nearby cool stars: GJ~9827, K2-18, TOI-1201, TOI-1685, and Barnard's Star. We selected these targets to span a broad range of effective temperatures ($T_{\rm eff} \in [3200,4250]$~K) and metallicities ([M/H] $\in [-0.4,+0.2]$), allowing us to test our methodology across different spectral types. 
GJ~9827, K2-18, TOI-1201, and TOI-1685 were specifically chosen, as they host transiting super-Earths or sub-Neptunes with planned or existing observations with either HST or JWST to characterize the planetary atmospheres. Having reliable stellar abundances will be essential to accurately interpreting the atmospheric compositions and planetary interiors in the context of the host star. We also include Barnard's Star, which serves as an important benchmark in the low $T_{\rm eff}$, metal-poor regime where abundance measurements are particularly difficult. Barnard's Star's elemental abundances have also been measured in multiple studies \citep{Maldonado2020,Ishikawa2022,Jahandar2024,Olander_2025}, providing several independent literature benchmarks for our abundance measurements. These systems are described in detail in the following sections and their adopted stellar parameters are compiled in Table~\ref{tab:stellarparams}.

\subsection{GJ~9827} \label{sec:GJ9827}
GJ~9827 is a nearby ($\sim$29.7 pc), bright ($J = 7.98$), moderately active K7 dwarf that represents the hottest star in our sample \citep{Prieto-Arranz2018, Passegger2024}. Observations with K2 revealed three transiting planets in the system: two inner super-Earths and an outer sub-Neptune, orbiting in a near 1:3:5 resonance chain \citep{Niraula_2017, Rodriguez_2018}. The planets around GJ~9827 span the radius valley, motivating several follow-up studies to characterize the planets and constrain formation scenarios \citep[e.g.][]{Teske_2018, Prieto-Arranz2018, Rice_2019}. HST/WFC3 spectra of GJ~9827~d revealed a tentative detection of water absorption \citep{Roy_2023}, while recent observations with JWST NIRISS/SOSS revealed a metal-rich atmosphere with O/H $\approx 4\times$ solar, providing evidence for a steam world scenario \citep{PiauletGhorayeb_2024}. Non-detections of Ly$\alpha$, H$\alpha$, and He~I cast doubt on an actively escaping envelope, favoring a long-lived, high mean molecular weight atmosphere \citep{Kasper_2020, Carleo_2021, Krishnamurthy_2023, PiauletGhorayeb_2024}. As one of the smallest exoplanets with a detected atmosphere, GJ~9827~d and the unique system architecture provide a valuable benchmark for studies of atmospheric evolution and composition across the radius valley.

For GJ~9827, we adopt the stellar parameters reported by \citet{Passegger2024}, who derived a $T_{\rm eff} = 4236\pm12$~K, and [Fe/H] = $-0.29\pm0.03$~dex, using a high S/N template spectrum obtained with ESPRESSO and analyzed with the SteParSyn spectral synthesis framework \citep{Tabernero_2022}. The resulting parameters are also in excellent agreement with previous determinations in the literature, such as \citet{Niraula_2017} ($T_{\rm eff} = 4255\pm110$~K, [Fe/H] = $-0.28\pm0.12$~dex), \citet{Rodriguez_2018} ($T_{\rm eff} = 4269\pm99$~K), \citet{Prieto-Arranz2018} ($T_{\rm eff} = 4219\pm70$~K, [Fe/H] = $-0.29\pm0.12$~dex), \citet{Rice_2019} ($T_{\rm eff} = 4340^{+48}_{-53}$~K, [Fe/H] = $-0.26\pm 0.09$~dex), and \citet{Kosiarek_2021} ($T_{\rm eff} = 4294\pm52$~K, [Fe/H] = $-0.26\pm0.08$~dex).

\subsection{K2-18} \label{sec:K2-18}
K2-18 is a moderately active M2.5 dwarf located $\sim$38~pc away that hosts a compact two-planet system consisting of an inner non-transiting super-Earth, K2-18~c \citep{Cloutier_2017,Cloutier_2019,Radica_2022}, and an outer transiting temperate sub-Neptune, K2-18~b \citep{ForemanMackey_2015, Montet_2015}. Owing to its location within the habitable zone and its favorable transmission spectroscopy geometry, K2-18~b has become one of the most extensively studied sub-Neptunes \citep{Montet_2015, Crossfield_2016, Benneke_2019, Tsiaras_2019}. Its measured mass and radius are consistent with either a volatile-rich interior or a rocky core surrounded by a substantial H$_2$/He envelope \citep{Cloutier_2019, Benneke_2019, Madhu2020_k218,LuquePalle2022,Rogers_2023, FernandezK218_2025}. While early HST observations revealed H$_2$O and subsequent JWST spectra confirmed CH$_4$, the planet's atmospheric composition and interior structure remain actively debated, with recent studies generally favoring a gas-rich mini-Neptune over a habitable Hycean world \citep{Madhu2023_DMS, Wogan_2024,Leconte_2024,Shorttle_2024,SchmidtMacDonald_2025,Luque_2025}. 

Given its well-characterized yet highly debated planetary atmosphere, together with the detailed stellar abundance analysis of \citet{Hejazi_2024}, the K2-18 system provides an important laboratory for exploring connections between host-star composition, planet formation pathways, and atmospheric evolution. The work of \citet{Hejazi_2024} also serves as the primary point of comparison with the abundances derived in this study. We therefore adopt the stellar parameters from \citet{Hejazi_2024}, who derived $T_{\rm eff} = 3547\pm85$~K and [Fe/H]$=0.17\pm0.10$~dex using high-resolution $H$- and $K$-band IGRINS spectra. Because our analysis is likewise based on high-resolution NIR spectroscopy and employs a similar spectral synthesis methodology, adopting the parameters of \citet{Hejazi_2024} provides the most direct comparison between the two analyses while also taking advantage of stellar parameters derived from wavelength regions particularly well-suited to characterizing M dwarfs. Other studies report somewhat different values using alternative techniques. For example, \citet{Sairam_2025} derived $T_{\rm eff} = 3645\pm52$~K and [Fe/H] $=0.10\pm0.12$~dex from pseudo–equivalent width measurements of Fe lines in optical HARPS spectra, while \citet{Hardegree_2020} estimated $T_{\rm eff} = 3590\pm98$~K and [Fe/H]$=0.26\pm0.17$~dex. Although these values are broadly consistent with one another, they illustrate the systematic differences that can arise when using different datasets and analysis methods. 

\subsection{TOI-1201} \label{sec:TOI1201}
TOI-1201 is a moderately young ($\sim$600--800 Myr) and bright (J $\approx$ 9.5 mag) early M dwarf located 37.9~pc away in a wide binary system with TOI-393 \citep{Kossakowski_2021}. A transiting sub-Neptune, TOI-1201~b, with an orbital period of 2.49 days was discovered, making it a relatively rare example of a planet orbiting an M dwarf in a multiple-star system \citep{Kossakowski_2021}. The young age of the system makes it an excellent candidate for studying mechanisms of atmospheric loss \citep{Kossakowski_2021}. TOI-1201~b is an especially favourable target for future atmospheric characterization, and was recently observed with the HST as part of the SPACE Program (Proposal GO-17192). 

Subsequent work revised the stellar and planetary properties, finding a smaller stellar radius ($0.481\pm0.014 R_\odot$ versus $0.508\pm0.016 R_\odot$) and consequently, a smaller planetary radius ($2.183\pm0.088R_\oplus$) \citep{Gore_2024}. The two studies also report significantly different stellar parameters, with \citet{Gore_2024} deriving $T_{\rm eff}=3659\pm73$~K and [Fe/H] $=-0.09\pm0.08$~dex, compared to $T_{\rm eff}=3476\pm51$~K and [Fe/H] $=0.05\pm0.16$~dex from \citet{Kossakowski_2021}. These discrepancies primarily reflect the use of different methodologies: \citet{Kossakowski_2021} derived the stellar parameters through spectral synthesis of CARMENES VIS spectra, whereas \citet{Gore_2024} employed empirical M-dwarf relations together with \textit{Gaia} photometry and NIR spectra to obtain largely model-independent stellar parameters. \citet{Marfil_2021}, using CARMENES VIS and NIR spectra analyzed with SteParSyn, derived $T_{\rm eff}=3655\pm25$~K, supporting the higher \teff\ reported by \citet{Gore_2024}. We therefore adopt the stellar parameters from \citet{Gore_2024} throughout this work. 

\subsection{TOI-1685} \label{sec:TOI1685}
TOI-1685 is an M3 dwarf ($J = 9.6$) located 37.6 pc away that hosts the ultra-short-period super-Earth TOI-1685~b \citep{Bluhm_2021, Hirano_2021, Burt_2024}. The planet was independently discovered by \citet{Bluhm_2021} and \citet{Hirano_2021}, whose differing stellar parameters and RV fit results led to conflicting interpretations of the planet as either volatile-rich or predominantly rocky. Subsequent RV characterization with CARMENES, IRD, and MAROON-X revealed an Earth-like density of 5.3\,g/cm$^3$, suggesting a rocky planet with little or no volatile envelope \citep{Burt_2024}. This picture has since been reinforced by JWST transmission and phase curve observations, which are consistent with a dark, bare-rock surface lacking a substantial atmosphere \citep{Luque_2024,Fisher_2026}. As an extremely short-period rocky planet, TOI-1685~b provides an opportunity to explore atmospheric loss near the cosmic shoreline, and investigate whether such a planet formed with no atmosphere, or if the atmosphere was stripped off in its lifespan.

The stellar parameters reported for TOI-1685 vary considerably across the literature due to differences in observational data and analyses. Using high-resolution CARMENES spectra and the spectral synthesis framework of \citet{Passegger_2019}, \citet{Bluhm_2021} derived $T_{\rm eff}=3434\pm51$~K and [Fe/H] $=-0.13\pm0.16$~dex. Applying an equivalent-width methodology to Subaru/IRD spectra, \citet{Hirano_2021} obtained a similar effective temperature ($3428\pm97$~K) but a higher metallicity ([Fe/H] $=0.14\pm0.12$~dex). Using APOGEE $H$-band spectra analyzed with BACCHUS, MARCS, and Turbospectrum, \citet{Melo2024} reported $T_{\rm eff}=3519\pm100$~K and [Fe/H] $=0.06\pm0.18$~dex. More recently, \citet{Burt_2024} estimated the stellar metallicity from photometric calibrations, finding an even higher value of [Fe/H] $\approx +0.3$~dex. We adopt the stellar parameters from \citet{Melo2024} to provide a consistent baseline for comparison with previously published abundance measurements.

\subsection{Barnard's Star} \label{sec:Barnard}
Barnard's Star (GJ 699) is one of the brightest ($J=5.2$) and closest M dwarfs at a distance of only 1.83 pc, making it an ideal benchmark star for M dwarf analyses in the literature. Although several earlier reports of planet detections around Barnard's Star \citep{vandeKamp_1963, Ribas2018} were later disproven \citep{Gatewood_1995, Benedict_1999, Lubin_2021, Artigau_2022}, the unprecedented RV precision of modern spectrographs such as ESPRESSO and MAROON-X has enabled the detection of much lower-mass companions for the first time. These observations have confirmed four non-transiting sub-Earth-mass planets around Barnard's Star, with orbital periods between 2.34 and 6.74 days \citep{Gonzalez-Hernadez_2024-Barnardplanet, Basant_2025}. While much too close to the star to be considered habitable, Barnard's Star and its planets represent a fascinating compact planetary system and provide a critical reference for investigating planet formation at the lowest masses.

Despite the extensive observational history of Barnard's Star, its stellar parameters have shown notable dispersion in the literature, with reported metallicities ranging from -0.86~dex \citep{Marfil_2021} to +0.61~dex \citep{Passegger_2022}. Reported effective temperatures also vary dramatically from 3092~K \citep{Hojjatpanah_2019} to 3463~K \citep{Foque_2018}. More recent spectroscopic analyses have converged toward a narrower range of parameters, and though they vary slightly depending on the analysis methodology, the following values are broadly consistent within uncertainties. \citet{Schweitzer_2019}, using CARMENES VIS spectra, derived $T_{\rm eff}=3273\pm51$~K and [Fe/H] $=-0.15\pm0.16$~dex. \citet{Marfil_2021}, using both CARMENES VIS and NIR data, reported $T_{\rm eff}=3254\pm32$~K, and [Fe/H] $=-0.57\pm0.10$~dex. \citet[][hereafter \citetalias{Jahandar2024}]{Jahandar2024} derived $T_{\rm eff}=3231\pm21$~K and [Fe/H] $=-0.39\pm0.03$~dex using SPIRou. \citet{Gonzalez-Hernadez_2024-Barnardplanet}, using the SteParSyn framework \citep{Tabernero_2022}, derived $T_{\rm eff}=3195\pm 28$~K and a [Fe/H] of $-0.56\pm0.07$~dex. For the purposes of this paper, we adopt the values from \citetalias{Jahandar2024}. Their methodology was shown to recover effective temperatures consistent with model-independent interferometric measurements, and the inclusion of the $K$-band, now recognized as important for correctly constraining \teff\ in cool M dwarfs, provides an additional advantage over results derived from optical or $YJH$ spectra \citepalias{Jahandar2024}.

\input{Tables/stellarparams}

\section{High-Resolution Near-Infrared Spectroscopy with SPIRou} \label{sec:spirou}
The SpectroPolarim\`etre InfraRouge (SPIRou) instrument is a stabilized, high-resolution, near-infrared (NIR), fiber-fed, \'echelle spectropolarimeter located at the 3.6~m Canada–France–Hawaii Telescope (CFHT) on Maunakea in Hawaii. SPIRou's spectral coverage spans the $YJHK$ bands from $0.97-2.49\, \mu$m with a spectral resolving power of $R\sim 70,000$ \citep{Donati2020}. While SPIRou's design is optimized to characterize exoplanets around low-mass stars using precision radial velocimetry at the $1-2$~m/s level and map stellar magnetic field topologies using spectropolarimetry, its spectral domain and resolving power make it an ideal instrument to study cool star chemistry. Unlike for FGK-type stars, the NIR is particularly advantageous for abundance studies of M dwarfs, whose optical spectra are intrinsically faint and dominated by dense forests of blended absorption features and broad molecular bands that obscure the continuum \citep{Souto_2017}. Furthermore, SPIRou is especially valuable among high-resolution NIR spectrographs as it covers the $K$-band, centred at $\sim 2.2\,\mu$m, thus providing access to dozens of CO lines that are essential for measuring [C/H] and C/O ratios. Without $K$-band access, [C/H] is difficult to constrain in cool stars because the available C I lines are few and shallow, making carbon measurements based solely on atomic features comparatively uncertain \citep[e.g.][]{Asplund_2021, Jahandar_2025}.

We observed GJ~9827, K2-18, TOI-1201, and TOI-1685 with SPIRou between August and December 2024 as part of the observing program CFHT24BC29 (PI: N. Gromek), with additional exposures for GJ~9827 and TOI-1685 supplemented from other programs. Our fifth target is Barnard's Star, whose spectra were taken as part of the SPIRou Legacy Survey (PI: J.-F. Donati) and whose co-added spectrum was published in \citetalias{Jahandar2024}. Barnard's Star served as a SPIRou RV standard, resulting in a very high signal-to-noise ratio of $\sim 1000$. The details of the observations are provided in Table~\ref{tab:observationstable}.

We reduced our data using A PipelinE to Reduce Observations \citep[APERO;][]{Cook2022}, SPIRou's data reduction software. APERO performs image calibration, wavelength solution, 1D spectral extraction, and telluric correction \citep{Artigau_2014}, yielding residual telluric contamination below 1\% of the continuum and enabling reliable high-precision RV and stellar parameter analysis \citep{Cook2022}.

\begin{deluxetable*}{lccccc}
\tablecaption{Summary of Observations}
\label{tab:observationstable}
\tablehead{
\colhead{Target} &
\colhead{$N_{\rm exp}$\tablenotemark{a}} &
\colhead{$t_{\rm exp}$ (s)\tablenotemark{b}} &
\colhead{Median S/N\tablenotemark{c}} &
\colhead{Observing Mode} &
\colhead{Source}
}
\startdata
GJ 9827 & 23 & 455.2 & 192.23 & OBJ\_FP & \makecell{CFHT24BC29 (PI: N. Gromek; $N_{\rm exp}=4$) \\ CFHT20BC27 (PI: J. Sikora; $N_{\rm exp}=19$)}  \\
K2-18 & 7 & 902.7 & 116.68 & OBJ\_FP & CFHT24BC29 (PI: N. Gromek) \\
TOI-1201 & 6 & 835.8 & 124.29 & OBJ\_FP & CFHT24BC29 (PI: N. Gromek)\\
TOI-1685 & 12 & 844.1 & 120.16 & OBJ\_FP & \makecell{
CFHT24BC29 (PI: N. Gromek; $N_{\rm exp}=7$) \\ CFHT23BC28 \& CFHT24BC22 (PI: R. Cloutier; $N_{\rm exp}=5$)} \\
Barnard's Star & 1373 & 60.0 & > 104.46 & POLAR\_FP & SPIRou Legacy Survey (PI: J.-F. Donati) \\
\enddata
\tablenotetext{a}{$N_{\rm exp}$ denotes the number of individual exposures.}
\tablenotetext{b}{$t_{\rm exp}$ is the mean exposure time of the individual spectra used in the co-addition, weighted by the number of exposures when observations from multiple programs are combined.}
\tablenotetext{c}{The median S/N per pixel among the individual exposures, measured near $1.635\, \mu$m, the central wavelength of order 33.}
\end{deluxetable*}

\section{Methodology for Measuring Stellar Elemental Abundances} \label{sec:methods}
Our elemental abundance framework is based on a spectral synthesis approach. Spectral synthesis is better suited to the complex spectra of M dwarfs than equivalent width methods, where there is extensive molecular absorption and significant line blending. Equivalent widths can provide inaccurate abundance estimates where lines are blended, saturated or otherwise deviate from Gaussian profiles, whereas spectral synthesis will model the full line profile \citep{Freckelton_2024}. Much of the functionality in our procedure is adapted from the open-source stellar spectroscopy code \texttt{iSpec} \citep{BlancoC2014,BlancoC2019}. \texttt{iSpec} serves as a wrapper combining a stellar atmosphere model with an appropriate atomic linelist and radiative transfer code to generate a synthetic spectrum given input stellar parameters, including individual elemental abundances [X/H]. We derive elemental abundances by fitting synthetic spectra to our order-merged, telluric-corrected SPIRou spectra on a line-by-line basis. Here, we describe our step-by-step procedure in detail.

\subsection{Global radial velocity shift and continuum normalization}
Each spectrum is shifted to the rest frame by applying a radial velocity (RV) correction. We determine this shift by cross-correlating the spectrum with a synthetic solar template generated from the same linelist used in our line-by-line analysis. We adopt the RV corresponding to the peak in the cross-correlation function and apply a global RV shift using the \texttt{ispec.correct\_velocity} function. We then perform a global continuum normalization using the functions \texttt{ispec.fit\_continuum} and \texttt{ispec.normalize\_spectrum}. We adopt the default median+max spline filtering scheme to fit the continuum across the entire spectrum. Spectral peaks are initially identified and suppressed using a maximum filter with a width of 1.0 nm, followed by a median filter of width 0.05 nm to further smooth the spectrum. After filtering, quadratic B-splines are fit to 5 nm wide segments across the full wavelength range to ensure a consistent continuum normalization.

\subsection{Linelist Determination}
Reliable abundance measurements require spectral features that are accurately identified and sufficiently free of blends. We therefore construct and vet a dedicated linelist for each target in our sample. Atomic and molecular transitions from CO, OH, and CN are taken from the Vienna Atomic Line Database \citep[VALD3;][]{Ryabchikova_2015}, while H$_2$O transitions are adopted from the BT2 linelist \citep{Barber_2006}. Specifically, the strongest $\sim1\%$, or 1,865,037, water transitions were included. Although we do not directly use H$_2$O features as abundance diagnostics, they contribute significantly to suppressing the pseudo-continuum and causing molecular blending in cool stars, and are included to ensure sufficient accuracy in the model spectra \citep{Allard2000, Tsuji_2015, Souto_2017}. Candidate absorption lines for our line-by-line analysis are initially selected from the shifted and normalized spectrum upon cross-referencing the spectrum with atomic and molecular transitions in the VALD3 linelist. This initial line selection is performed using \texttt{ispec.find\_linemasks} according to the following criteria:

\begin{itemize}
    \item[] (i) The line depth must reach at least 10\% below the estimated continuum.
    \item[] (ii) The line maximum cannot extend beyond 100\% below the estimated continuum.
    \item[] (iii) The observed central wavelength must be within $\pm 0.015$ nm of the central reference wavelength in the VALD3 linelist when applied to SPIRou spectra. This value is adjusted as needed for the spectrograph used. 
\end{itemize}

This process produces an initial set of line masks that define the mask's central wavelength, with the edges defined by the nearest local flux maxima, as well as the corresponding chemical species. We proceed by refining this preliminary linelist through a two-stage vetting process to eliminate misidentified or severely blended features from which [X/H] values cannot be confidently recovered. We then perform an automated filtering step on each candidate line mask whereby we first generate a grid of seven synthetic spectra with [X/H] spanning $\pm0.75$~dex in increments of 0.25~dex. We remove line masks for which no corresponding synthetic feature is produced, as well as lines whose fits converge beyond our [X/H] grid. We note that NIR linelists are known to be incomplete, particularly for very cool stars \citepalias{Jahandar2024}, with some visibly prominent lines present in the observed spectrum that are not replicated in our linelist. Consequently, these features are not included in our final linelist.

\begin{figure*}
    \centering
    \includegraphics[width=\linewidth]{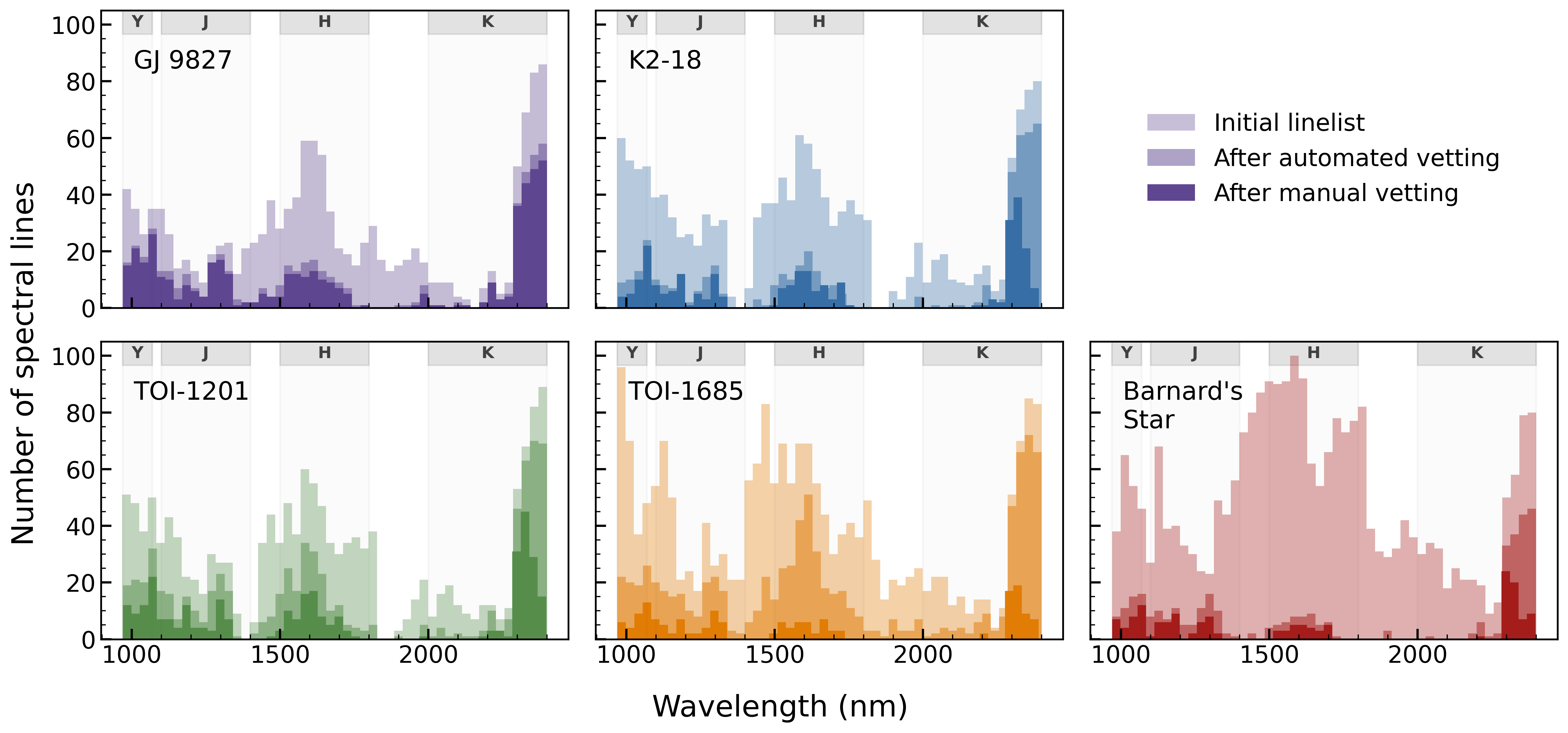}   
    \caption{Wavelength distributions of candidate spectral lines retained at each step of the line-selection procedure. The histograms show the initial linelist (light), lines retained after automated vetting (intermediate), and the final manually vetted sample (dark). The shaded background regions indicate approximate wavelength ranges for the $Y, J, H,$ and $K$-bands.}
    \label{fig:LineDensityPlot}
\end{figure*}

Following the automated selection, we visually inspect the vetted lines and exclude those affected by blending, residual telluric contamination, or spectral reduction artifacts, all of which result in mismatches between the model and observed spectra that skew the derived abundances and lead to inaccurate results. These model discrepancies are especially pronounced for cool M dwarfs where molecular absorption bands (e.g. FeH, H$_2$O, etc.) can significantly contaminate atomic absorption lines \citep{Allard2000, Tsuji_2015}. Our final linelist therefore retains only the lines that are reliably reproduced in our spectral synthesis models, biasing the sample to lines that are largely free of significant contamination. The number of spectral lines saved in each vetting step is visualized in Figure \ref{fig:LineDensityPlot}, and the final linelists used for our target stars are reported in Appendix~\ref{sec:full_linelist}. The wavelength distribution of selected features is highly non-uniform, primarily due to the strong telluric absorption between the $H$ and $K$-bands. Conversely, the sharp peak in the $K$-band is dominated by the dense forest of CO absorption features located there. The number of features in the final linelist also shows a broad dependence on \teff. Even though the cooler M dwarf spectra have line-rich spectra at first glance, the increased molecular absorption and blending make it more difficult to identify clean, isolated atomic features, thus reducing the number of retained lines. This effect is especially apparent when comparing the final linelists for GJ~9827 (\teff\ $= 4236$~K, 470 lines) and Barnard's Star (\teff\ $= 3231$~K, 160 lines).

While atomic features are used for most elements in our abundance analyses, the abundances of carbon and oxygen are derived from CO and OH molecular features, respectively. Atomic C I and O I features are known to be too shallow and are comparatively sparse compared to the rich number of CO and OH features in this temperature regime, thereby making atomic carbon and oxygen lines generally unreliable for our abundance measurements \citep[e.g.][]{Souto_2017}. The use of CO and OH lines to measure [C/H] and [O/H] is in line with many abundance studies of cool stars in the literature \citep[e.g.][]{Souto_2017,Hejazi2023,Hejazi_2024,Sharma_2024}. This choice is well-motivated at NIR wavelengths, which contain a wealth of OH lines in the $H$-band and CO lines in the $K$-band. Access to these bands with SPIRou significantly increases the number of CO and OH lines over C I and O I features, thus improving our measurement statistics for these elements. We do not use H$_2$O lines to measure [O/H] as they are less reliably replicated in spectral synthesis compared to OH and are known to exhibit a strong systematic dependence on \teff{} \citep{Souto2020,Jahandar2024}. Small uncertainties in \teff{} would therefore lead to large systematic errors in the inferred oxygen abundance when using H$_2$O lines, making OH lines a more robust diagnostic \citepalias{Jahandar2024}.

\subsection{Stellar Parameters}
Our spectral synthesis requires input stellar parameters, such as the stellar effective temperature (\teff{)}, overall metallicity ([M/H]), and surface gravity ($\log{g}$). As discussed for each of our targets in Section~\ref{sec:stars}, the stellar parameters \teff{,} [M/H], and $\log{g}$ can suffer significant systematics across methods and datasets. As our framework has not been rigorously tested for the determination of stellar parameters other than [X/H], we elect to adopt parameter values \teff{,} [M/H], and $\log{g}$ from the literature, looking for the confluence of the parameter values from multiple studies to inform our selection. In the final step of our abundance analysis, we quantify the errors introduced by varying the input stellar parameters. For applications of the framework in which literature constraints for $\log{g}$ are unavailable, we derive the stellar mass $M_\star$ and radius $R_\star$, and hence $\log{g}$, from the empirical $M_\star-M_{K_s}$ and $R_\star-M_{K_s}$ relations from \citet{Mann2015} and \citet{Mann_2019}, respectively. Similarly for \teff{}, we use the \teff{--}$(J-H)$--$(BP-RP)$ color relation from \cite{Mann2015}. These calculations rely on 2MASS $JHK_s$ photometry and $BP,RP$ photometry and parallaxes from Gaia DR3 \citep{Gaia_DR3}. Table~\ref{tab:stellarparams} summarizes our adopted stellar parameters and sources for each target.

Within our framework, spectral synthesis models uniformly scale the abundances of all elements according to the input stellar metallicity [M/H]. Regarding our adopted [M/H] values, we note that many spectroscopic studies of our targets use [Fe/H] as a proxy for [M/H]. This may be seen as a natural choice given that Fe lines are often numerous and readily available in visible-NIR spectra of cool stars. While we prioritize measurements of [M/H] when available, in cases where only [Fe/H] is available, we follow the convention of using [Fe/H] as a proxy for [M/H]. We do so while acknowledging that this is not always a reliable assumption given that in cool stars, particularly metal-poor stars, the overall metal content can differ significantly from [Fe/H] due to an increased $\alpha$-enhancement \citep{Shan_2021,Marfil_2021}.

\begin{figure}
    \centering
    \includegraphics[width=\linewidth]{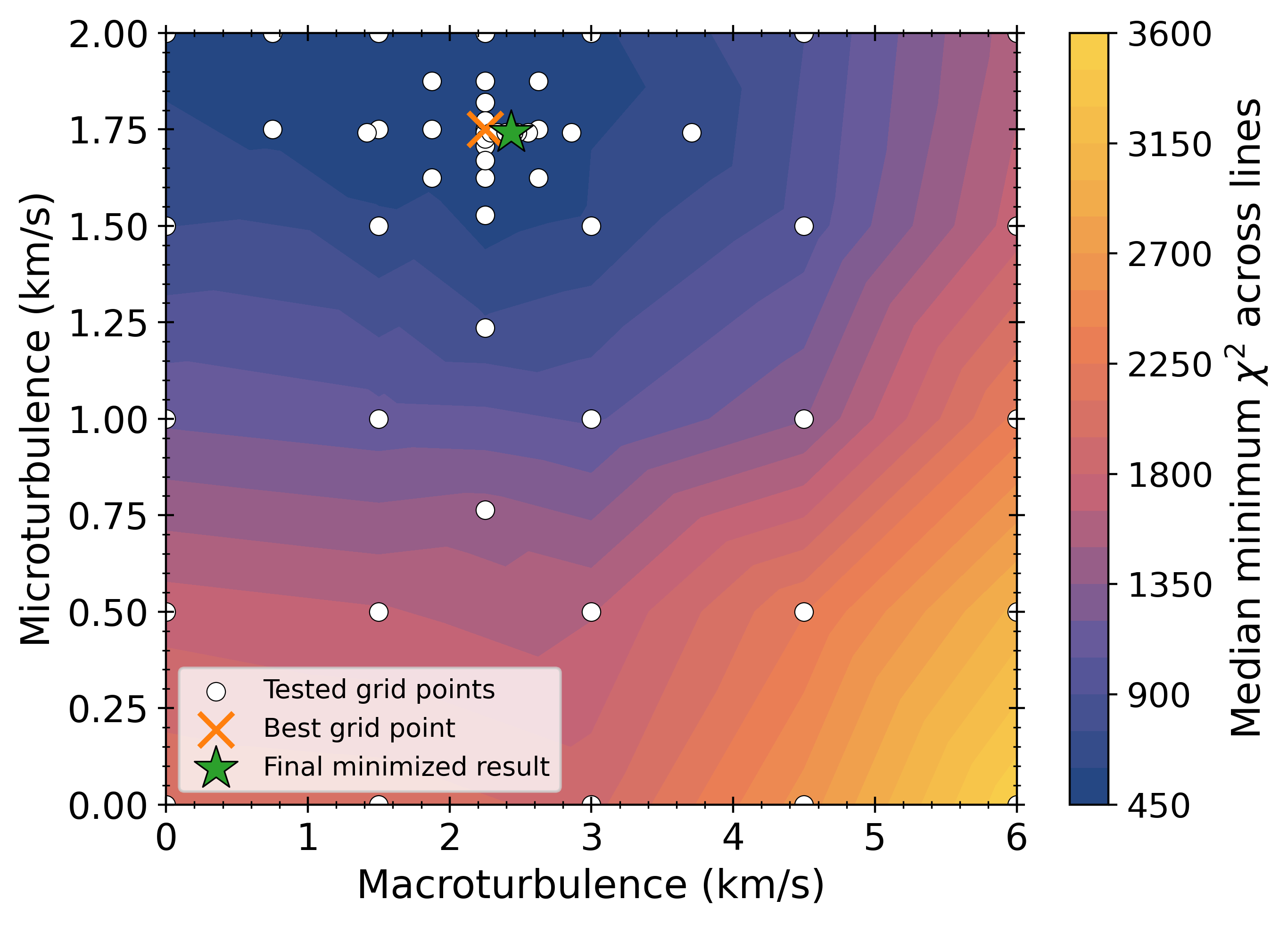}
    \caption{Determination of $v_{\rm mic}$ and $v_{\rm mac}$ for TOI-1201. The filled contours show the median minimum $\chi^2$ measured across the selected spectral lines as a function of $v_{\rm mac}$ and $v_{\rm mic}$. White circles indicate combinations of $v_{\rm mac}$ and $v_{\rm mic}$ explicitly evaluated during the grid search. The orange cross marks the lowest-$\chi^2$ point found on this discrete grid, while the green star shows the final minimized solution.}
    \label{fig:macnmicroturb}
\end{figure}

\subsubsection{Line-Broadening Parameters: Macroturbulent, Microturbulent, \& Rotational Velocities} \label{sec: broadparams} 
Our synthesis additionally requires prescriptions for microturbulent and macroturbulent broadening, in order to approximate unresolved convective and granulation effects \citep{Asplund_2000, BrewerFischerCoolStars2016, BrewerFischerMgSiCO2016}. Microturbulence ($v_{\rm mic}$) is incorporated by Turbospectrum during the radiative-transfer calculation and therefore affects line formation and strength, whereas macroturbulence ($v_{\rm mac}$) is applied by \texttt{iSpec} as a radial-tangential convolution of the synthesized spectrum, symmetrically broadening the lines without affecting their strength \citep{BlancoC2014, BrewerFischerCoolStars2016, BlancoC2019}. As such, both parameters have a significant impact on spectral line shape, and influence the inferred abundances. Within \texttt{iSpec}, empirical relations exist for $v_{\rm mac}$ and $v_{\rm mic}$ as a function of \teff{,} $\log{g}$, and [M/H]; however, these are calibrated from observations of FGK stars and are therefore not reliable for the cool stars considered in this study \citep{Ramirez_2013, Doyle_2014}. We instead follow a procedure similar to that of \citet{Hejazi2023} to estimate these nuisance parameters, which we do not claim to be measured values. 

We estimate $v_{\rm mac}$ and $v_{\rm mic}$ using molecular features, as they are significantly more sensitive to changes in broadening parameters than atomic features \citep{Souto_2017, Hejazi2023}. CO lines in the $K$-band are used when available. When using spectrographs that lack access to the $K$-band, OH lines can be used instead. We generate synthetic spectra over a two-dimensional grid spanning $v_{\rm mac} \in [0,6]$~km/s and $v_{\rm mic} \in [0,2]$~km/s in steps of 1~km/s and 0.25~km/s, respectively. Our $v_{\rm mic}$ and $v_{\rm mac}$ parameter range encompasses values expected for cool stars, with reported values for M dwarfs of $v_{\rm mic}\sim 0.5 - 1.5$\,km/s \citep{Bean2006, Souto_2017, Hejazi2023} and $v_{\rm mac}\sim 0 - 3$\,km/s \citep{Bean2006, Lindgren2017, Cristofari_2022}, while allowing sufficient margin to accommodate star-to-star variations. Broadening due to stellar rotation and macroturbulent velocity will produce degenerate effects on line profiles, even at high spectral resolving power, thus making it difficult to reliably disentangle the two effects \citep{BrewerFischerCoolStars2016}. Consequently, we fix the sky-projected stellar rotational velocity ($v\sin{i_\star}$) to 0~km/s throughout our analyses. Our derived $v_{\rm mac}$ value should therefore be interpreted as an effective broadening term that incorporates both rotational and macroturbulent contributions \citep{BrewerFischerCoolStars2016, Shan_2021, Cristofari_2022}. This decision also motivated our definition of the range of $v_{\rm mac}$ between 0-6~km/s; not because such large macroturbulent velocities are expected for cool stars, but because $v_{\rm mac}$ is treated as a generalized broadening parameter.

At each grid point, we perform our spectral synthesis given $\{ v_{\rm mac},v_{\rm mic} \}$, fixed values of \teff{,} $\log{g}$, and [M/H] from Table~\ref{tab:stellarparams}, and by setting [C/H]$=$[M/H]. We compute the $\chi^2$ goodness-of-fit statistic at each grid point by comparing the synthetic and observed spectra for each CO line in the star's linelist. We then compute the median $\chi^2$ value across all lines to assess the overall quality of fit for each $\{ v_{\rm mac}, v_{\rm mic} \}$ combination. Taking the median reduces the influence of individual CO lines that may be affected by unidentified blends or weak sensitivity to the turbulent velocity parameters. The minimum $\chi^2$ identified from our coarse grid is subsequently refined with the derivative-free Powell algorithm, as implemented in \texttt{scipy.optimize.minimize} \citep{Virtanen2020}. The optimizer iteratively searches the local parameter space to determine the best-fit values of $v_{\rm mac}$ and $v_{\rm mic}$ that minimize $\chi^2$. This produces a two-dimensional $\chi^2$ surface describing the goodness-of-fit as a function of $v_{\rm mac}$ and $v_{\rm mic}$, as demonstrated in Figure~\ref{fig:macnmicroturb} for TOI-1201. We estimate uncertainties in $v_{\rm mac}$ and $v_{\rm mic}$ from the full width at half maximum (FWHM) of the one-dimensional $\chi^2$ profiles obtained by varying one turbulent velocity parameter while fixing the other at its best-fit value. Half of the measured FWHM is adopted as the empirical uncertainty, which we later propagate through to our uncertainties on [X/H] values. Our derived turbulent velocities are reported in Table~\ref{tab:stellarparams} and all remain within the expected ranges for M dwarfs.


\subsection{Abundance Analysis} \label{sec: abund_analysis_methods} 

We determine elemental abundances using spectral synthesis methods. We generate synthetic spectra using the one-dimensional, plane-parallel MARCS model atmospheres \citep{Gustafsson2008}, assuming local thermodynamic equilibrium (LTE), together with the radiative transfer code Turbospectrum v19.1 \citep{Plez2012}. We adopt solar abundances from \citet{Asplund2009}. 

For each spectral feature in our refined linelist, we first generate a grid of synthetic spectra that varies the corresponding elemental abundance $[\textrm{X/H}]\in [-0.75,+0.75]$ dex in steps of 0.25~dex, using the function \texttt{ispec.generate\_spectrum}. All other stellar parameters are initially held fixed to the values reported in Table~\ref{tab:stellarparams}. While the spectral synthesis is initially restricted to the aforementioned coarse abundance grid, we then linearly interpolate the model spectra onto a finer grid with a spacing of 0.015~dex using \texttt{scipy.interpolate.interp1d}. Because the synthetic line profiles vary smoothly and monotonically over this abundance interval, linear interpolation provides a computationally inexpensive approximation. Furthermore, the errors introduced to the final abundance measurements are negligible compared to the other error terms introduced by varying the stellar parameters. Each synthetic spectrum is generated over the pre-defined line mask, plus margins of 0.5 nm on either side of the mask, henceforth referred to as the continuum region. This extended region serves to provide additional wavelength coverage needed to characterize the local continuum and to identify potential contamination from nearby spectral lines. Each synthetic spectrum is convolved to match the resolution of the observations (R = 70,000).  

\begin{figure*}
    \centering
    \includegraphics[width=\linewidth]{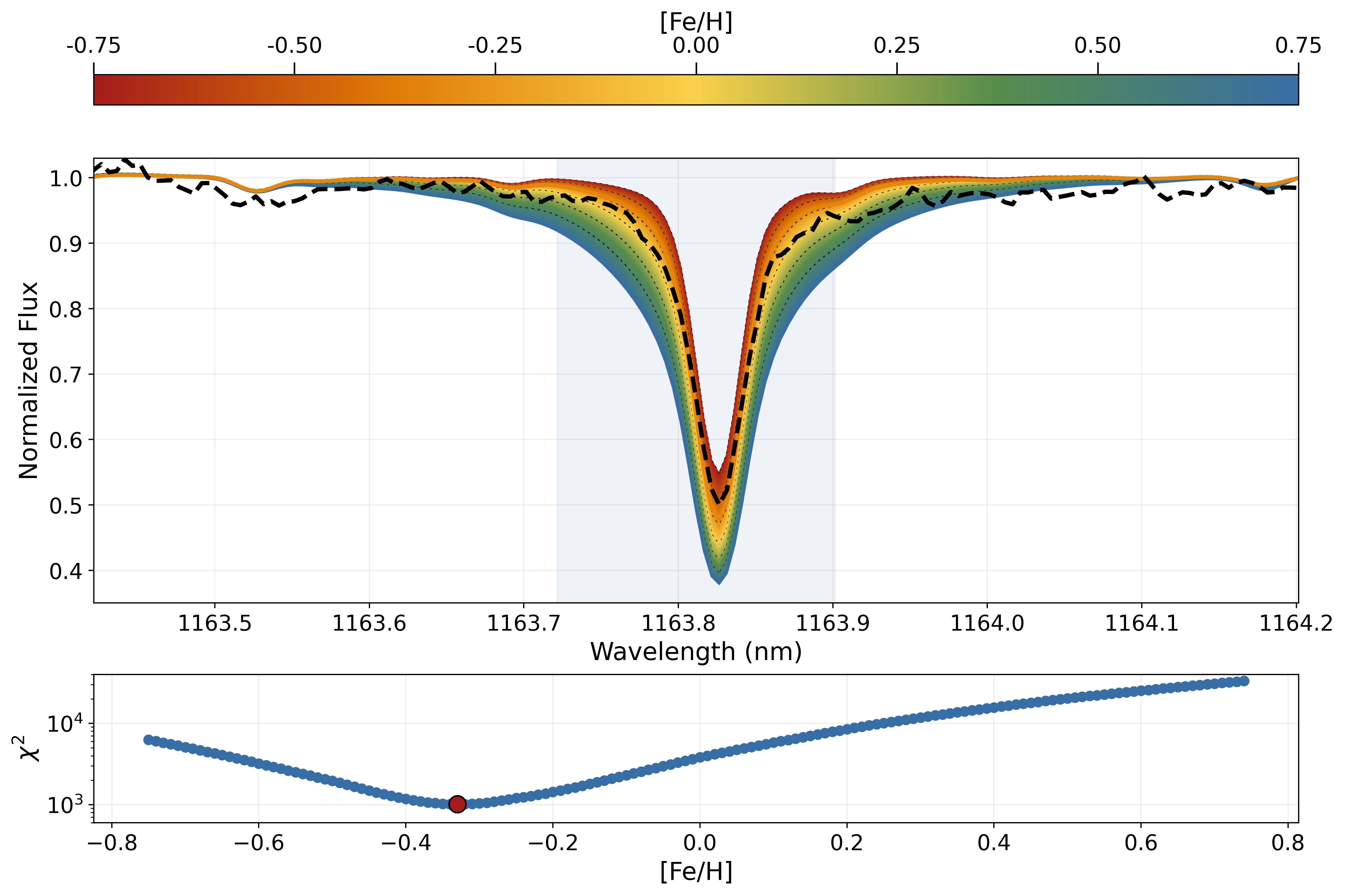}
    \caption{Sample determination of the Fe abundance from an individual Fe line at 1163.83 nm. Top: the observed normalized spectrum (black dashed) compared with the interpolated synthetic spectra spanning a range of [Fe/H] values, coloured according to abundance. The grey lines represent the initial synthesized spectra, before interpolation. The shaded blue region marks the line mask used for the $\chi^2$ minimization. Bottom: $\chi^2$ as a function of [Fe/H] for the same spectral line, with the red point marking the best-fit abundance.}
    \label{fig:SampleFeline}
\end{figure*}

We determine the best-fit abundance for each spectral line via $\chi^2$ minimization between the observed spectrum and the grid of model spectra. We calculate the $\chi^2$ statistic as a function of input [X/H] over the line mask after applying a wavelength shift (see Section~\ref{sec:wavelength_shift}) and normalization by the local pseudo-continuum (see Section~\ref{sec:PCN}). Figure~\ref{fig:SampleFeline} depicts an example of this process. The grid of synthetic spectra is shown compared to an observed Fe line in the spectrum of TOI-1201. The lower panel depicts the $\chi^2$ values as a function of [Fe/H] in the grid, calculated over the line mask. We repeat this step for every spectral line of a given element in our linelist. The final abundance for that element is calculated via a weighted average of the individual abundances obtained from the individual spectral lines. We calculate the weight for each line as the normalized line depth divided by the root-mean-square error (RMSE) between the best-fit synthetic model and the observed line in the line region. In this way, strong lines that are well reproduced by spectral synthesis contribute more weight to the final abundance measurement.

\subsubsection{Wavelength Alignment} \label{sec:wavelength_shift}
Even after applying a global RV correction, we find that many observed lines exhibit small wavelength offsets relative to the linelist rest frame (similarly reported in \citet{Jahandar2024}, see their Sec. 3.1.3). We align the spectrum on a line-by-line basis by applying a local wavelength shift prior to abundance determination. We identify the wavelength corresponding to the minimum flux of the observed feature and shift it to align with the wavelength corresponding to the minimum flux of the synthetic spectral line. To guard against potentially incorrect alignments in blended lines or crowded regions, this correction is not applied if a secondary minimum is detected within 10 resolution elements of the absolute minimum for the observed spectrum.

\begin{figure*}
    \centering

    \includegraphics[width=\linewidth]{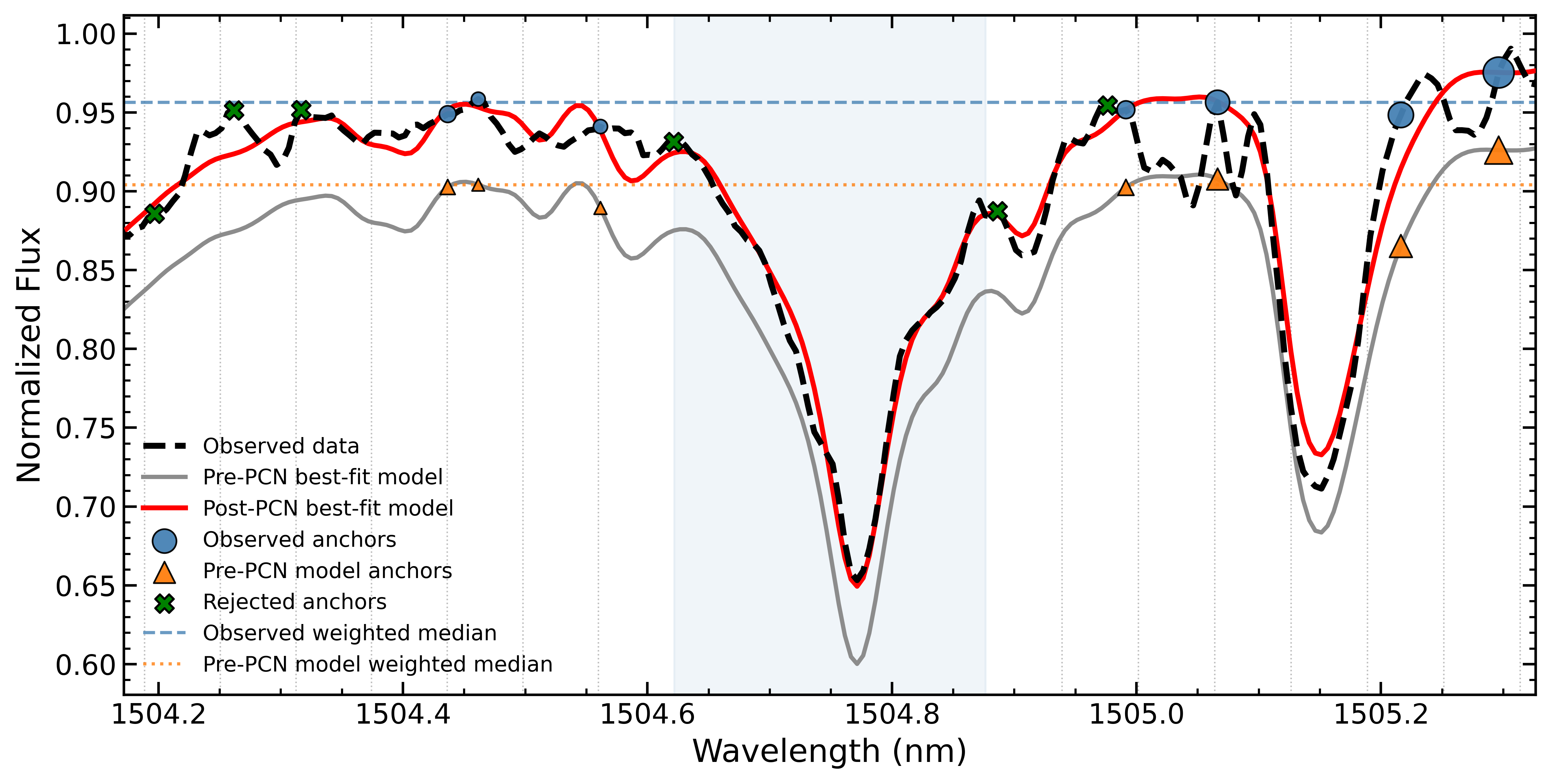}
    \caption{Visualization of the local pseudo-continuum normalization (PCN) procedure for an Mg feature at 1504.77 nm in TOI-1201. The observed spectrum (black dashed) is shown with the best-fitting model before (grey solid) and after (red) the continuum correction. The shaded region represents the line mask. Blue circles and orange triangles show the observed and model continuum anchors, respectively, while green crosses indicate rejected anchors. Marker sizes reflect the relative weights assigned to the accepted continuum points. The horizontal blue and orange lines show the weighted median levels of the observed and pre-PCN model anchor points, respectively. The vertical grey lines denote the segments used in the PCN process.}
    \label{fig:PCNimage}
\end{figure*}

\subsubsection{Pseudo-continuum Normalization} \label{sec:PCN}
The presence of molecular bands (e.g. H$_2$O and TiO) inhibits the identification of the true continuum in cool star spectra and necessitates the application of a local pseudo-continuum normalization (PCN) to avoid biasing [X/H] in a $\chi^2$-based framework. While there is general agreement on the need for a PCN to be implemented when analyzing spectroscopic abundances of cool stars, details of the PCN procedure can vary widely \citep{BlancoC2014, Hejazi2023, Melo2024, Medan_2025}, thus introducing variability and subjectivity \citep{Casey_2026}. Achieving an accurate and consistent placement of the pseudo-continuum is especially critical, as broadened lines, saturated lines or lines in regions with heavy contamination from molecular bands can all alter the apparent depths of the fitted lines and consequently bias the inferred abundances when left untreated  \citep{Souto_2017, Hejazi_2024}.

We apply a local PCN correction to each spectral feature. We define spectral windows spanning $\pm0.5$~nm around the line mask, hereafter referred to collectively as the continuum region. Data points within the central line mask are excluded from the PCN fitting. We divide the continuum region into sixteen equal segments to sufficiently sample variations in the local pseudo-continuum. We then flag the data point with the largest observed flux within each segment as a pseudo-continuum point. This choice adequately samples the upper envelope surrounding each line while mitigating the influence of nearby spectral lines that may overwhelm the PCN. We reject any continuum points where the highest-abundance model and observed line differ by more than 10\% in normalized flux. This both reduces the influence of emission lines or high intensity outliers, and removes any continuum points affected by the line wings in broad features. If only one out of 16 continuum points passes this criterion, the two points with the lowest difference in flux between model and observed data are retained to ensure that at least three points are used to define the continuum. At the wavelengths corresponding to each continuum point, we calculate the difference between the synthetic fluxes and the observed spectrum. For these values we calculate an additive offset from the weighted median of the observed-synthetic differences. The continuum points are individually weighted according to their agreement with the lowest-abundance model, such that points least affected by any abundance-sensitive absorption are most highly weighted. We apply the flux offset to each model spectrum prior to abundance determination. An example of this procedure is shown in Figure~\ref{fig:PCNimage}. 

By using an additive offset in place of a multiplicative normalization, as is common in the literature, we preserve the line profile in each synthetic spectrum. By anchoring the PCN to the highest-flux points in the continuum region, we reduce our sensitivity to line blending and flaws in the global continuum normalization without distorting the line profiles. For the vast majority of spectral lines, the shift caused by the PCN is small (i.e. $\lesssim1$\% in normalized flux), while lines with significant broadening are those most impacted by the PCN. While our PCN process is automated, we perform a visual inspection of the normalized lines to identify potential mismatches between the shifted synthetic spectra and the observed data in the continuum region, as this may be indicative of a flawed PCN. In such cases, the lines are removed as part of our linelist refinement process.

\subsection{Error Analysis}\label{sec:errors}
Uncertainties in the derived abundances arise from both line-by-line scatter and systematic uncertainties as a function of input stellar parameters. Here we quantify the individual error terms for the random and systematic contributions to the total abundance errors.  We calculate the uncertainty in [X/H] due to the dispersion across lines of the same species ($\sigma_{\rm ran}$) as the standard error of the mean of the [X/H] distribution containing $N$ lines (i.e. $\textrm{std}([\textrm{X/H}])/\sqrt{N}$). 

\begin{figure*}
    \centering
    \includegraphics[width=\linewidth]{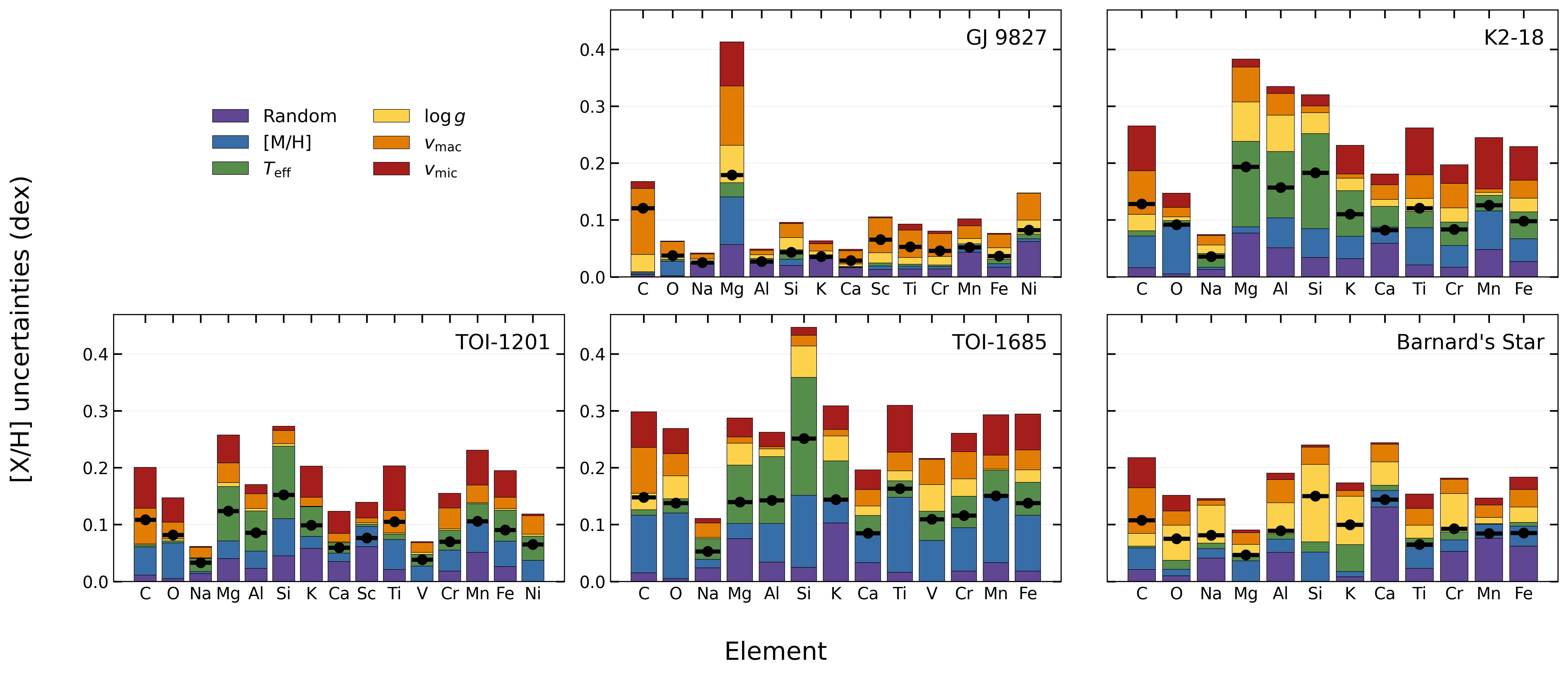}
    \caption{Total abundance uncertainties and their individual error contributions for each element and target star. The black lines represent the total uncertainty in [X/H], calculated by adding the individual uncertainty terms in quadrature. The coloured segments show the individual error terms, including the random line-to-line uncertainty and systematic contributions from [M/H], \teff, $\log g$, $v_{\rm mac}$, and $v_{\rm mic}$. These error terms are stacked for visualization purposes, and the total height of the bars does not represent the total error.}
    \label{fig:errorplot}
\end{figure*}

The systematic uncertainties associated with our adopted stellar parameters are calculated using a parameter perturbation analysis. For each parameter ($\{ T_{\rm eff}, \textrm{[M/H]}, \log{g}, v_{\rm mac}, v_{\rm mic} \}$), we resample 15 values from a Gaussian centered on its adopted value with a standard deviation equal to its $1\sigma$ uncertainty, while holding all other parameters fixed, and repeat the complete abundance analysis. The corresponding systematic uncertainty is the standard deviation of the resulting abundance distribution. We note that this treatment assumes that the stellar parameter uncertainties are independent and therefore does not include covariance between input parameters. Preliminary tests showed that these uncertainties typically converge after 6-12 realizations. The total abundance uncertainty is obtained by adding the line-by-line scatter and individual systematic contributions in quadrature. In this way, we are able to assess the contribution of each error term to the overall error budget of our abundance measurements. A table reporting the complete error analysis for each of our target stars is included in the Appendix (Table~\ref{tab:Error_table}), and a visualization of these errors for our stars is shown in Figure~\ref{fig:errorplot}.

\begin{figure*}
    \centering
    \includegraphics[width=\linewidth]{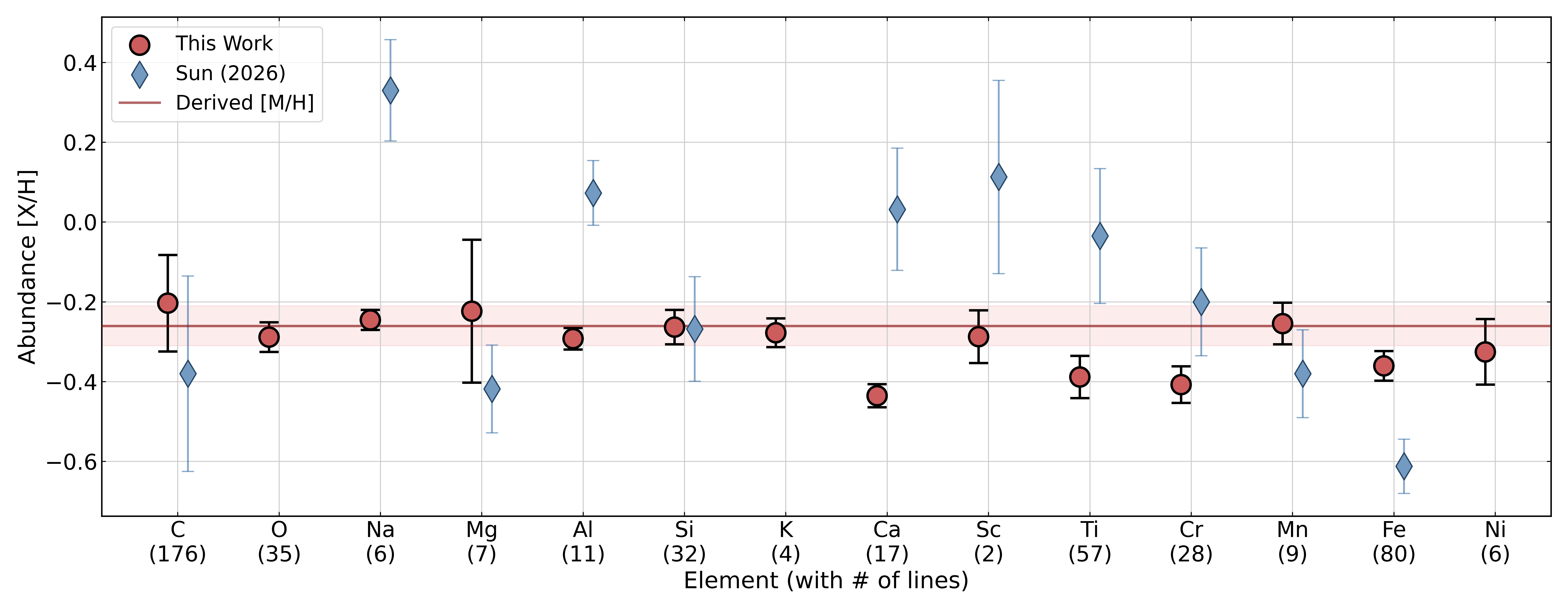}
    \caption{Elemental abundances of GJ~9827 derived in this work (red circles) compared with literature measurements from \citet{Sun_2026} (blue diamonds). The numbers below each element indicate the number of spectral lines used in this work. The horizontal line and shaded region show the derived [M/H] and its uncertainty.}
    \label{fig:abun_plots-gj9827}
\end{figure*}

\begin{figure*}
    \centering
    \includegraphics[width=\linewidth]{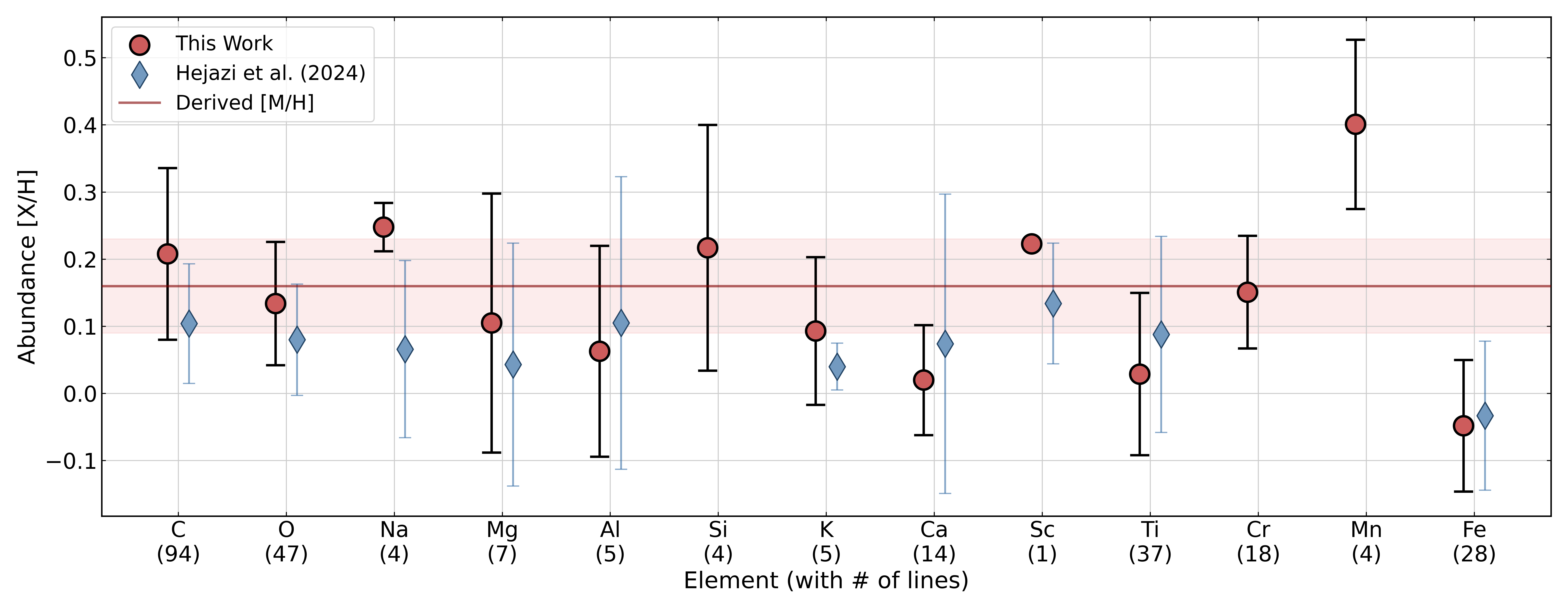}
    \caption{Elemental abundances of K2-18 derived in this work (red circles) compared with literature measurements from \citet[][blue diamonds]{Hejazi_2024}. The numbers below each element indicate the number of spectral lines used in this work. The horizontal line and shaded region show the derived [M/H] and its uncertainty.}
    \label{fig:abun_plots-k2-18}
\end{figure*}

\begin{figure*}
    \centering
    \includegraphics[width=\linewidth]{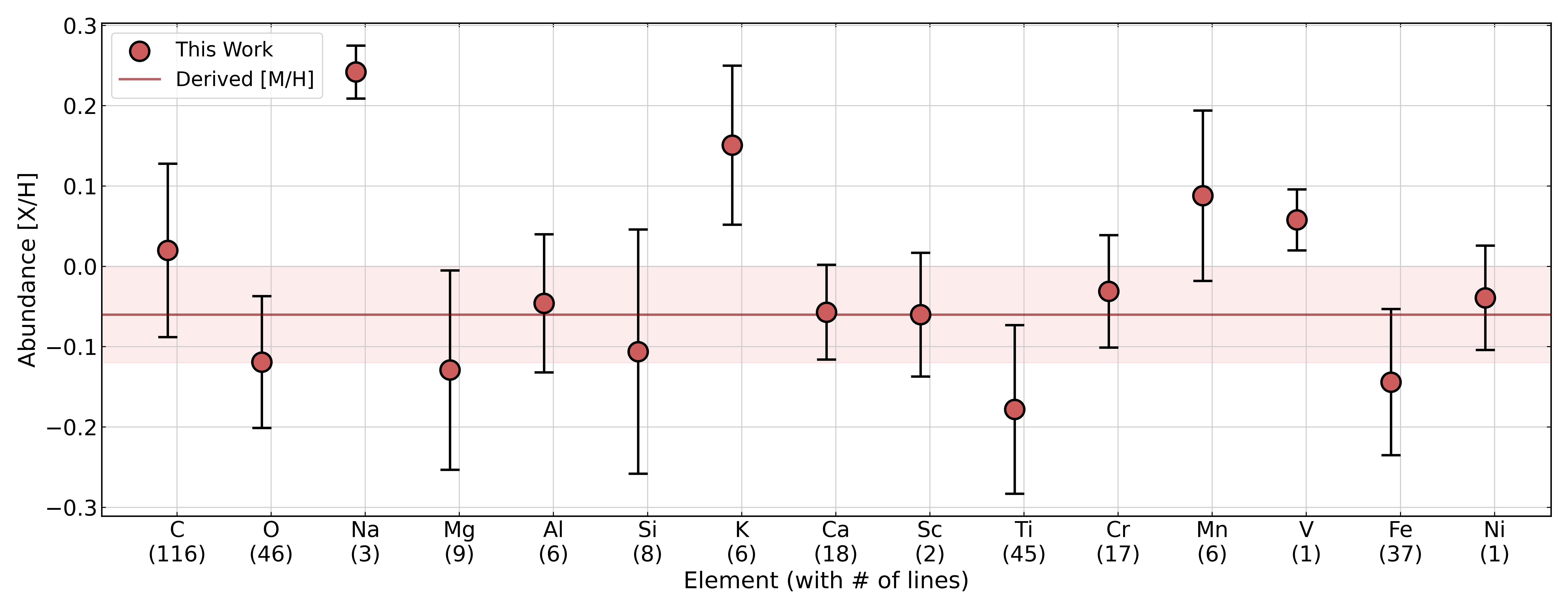}
    \caption{Elemental abundances of TOI-1201 derived in this work (red circles). The numbers below each element indicate the number of spectral lines used in this work. The horizontal line and shaded region show the derived [M/H] and its uncertainty.}
    \label{fig:abun_plots-toi1201}
\end{figure*}

\begin{figure*}
    \centering
    \includegraphics[width=\linewidth]{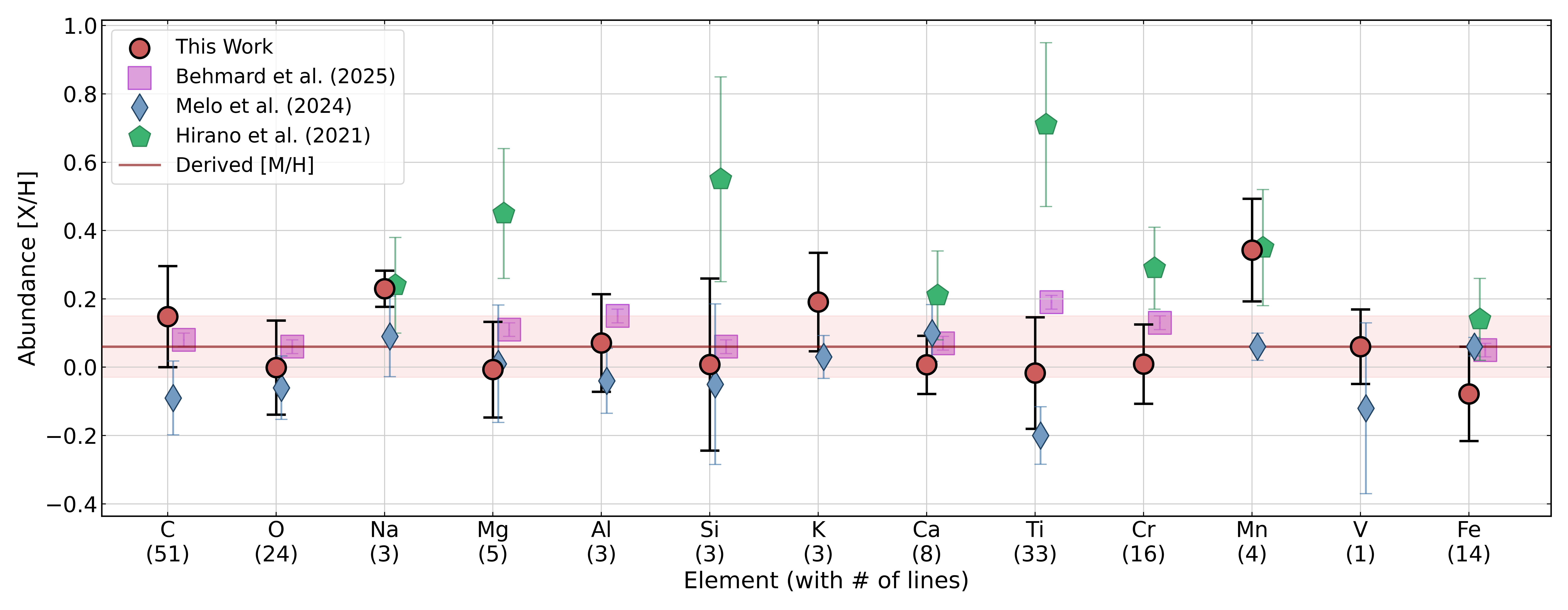}
    \caption{Elemental abundances of TOI-1685 derived in this work (red circles) compared with literature measurements from \citet[][green pentagons]{Hirano_2021}, \citet[][blue diamonds]{Melo2024}, and \citet[][purple squares]{Behmard_2025}. The numbers below each element indicate the number of spectral lines used in this work. The horizontal line and shaded region show the derived [M/H] and its uncertainty.}
    \label{fig:abun_plots-toi1685}
\end{figure*}

\begin{figure*}
    \centering
    \includegraphics[width=\linewidth]{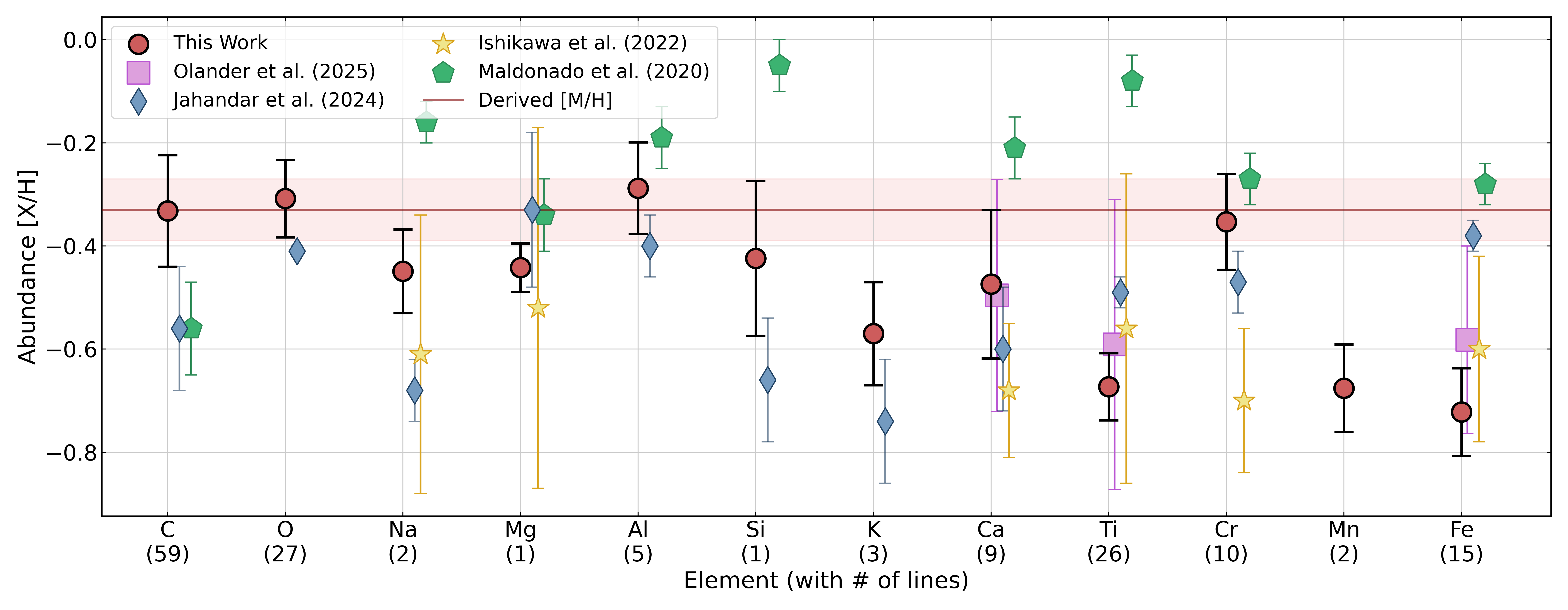}   
    \caption{Elemental abundances of Barnard's Star derived in this work (red circles) compared with literature measurements from \citet[][green pentagons]{Maldonado2020}, \citet[][yellow stars]{Ishikawa2022}, \citet[][blue diamonds]{Jahandar2024}, and \citet[][purple squares]{Olander_2025}. The numbers below each element indicate the number of spectral lines used in this work. The horizontal line and shaded region show the derived [M/H] and its uncertainty.}
    \label{fig:Barnardplot}
\end{figure*}

\section{Results} \label{sec:results}
We derive abundances of twelve elements (C, O, Na, Mg, Al, Si, K, Ca, Ti, Cr, Mn, Fe) for all five target stars, plus Sc, V, and Ni for a subset of stars. Our measured abundances are listed in Table~\ref{tab:abundance_table} and depicted in Figures~\ref{fig:abun_plots-gj9827}-\ref{fig:Barnardplot}. These values are the weighted averages of the line-by-line abundances for each element, according to the weighting scheme described in Section~\ref{sec: abund_analysis_methods}. While Table~\ref{tab:abundance_table} and Figures~\ref{fig:abun_plots-gj9827}-\ref{fig:Barnardplot} include the total combined errors on each [X/H], we provide complete error budgets for each element and each star in Table~\ref{tab:Error_table}. The complete linelists of all the atomic and molecular features used for each star are provided in Table~\ref{tab:complete_lines}. We note that there are some elements that are represented by only a small number of lines in select stars. Therefore, abundance determinations based on one or a few lines, such as V or Sc, should be treated with caution \citep{Adibek_2015}. We note that while our methodology in Section \ref{sec: abund_analysis_methods} lists the abundance grid bounds as $[-0.75,+0.75]$ dex, we extend these bounds to be $[-1.25,+1.25]$ dex for Barnard's Star only, to ensure the distribution of line-by-line abundances is not biased or limited by the boundaries. In this case, we keep the spacing of 0.25~dex and generate 11 synthetic models for interpolation.

From our measured elemental abundances we calculate bulk metallicities [M/H] following the formalism outlined in \citet{Hinkel2022}. The corresponding values are accurate representations of the definition of metallicity because they account for each element's absolute abundance. Recall that metallicity expressed as [M/H] is the molar fraction of elements heavier than He, here expressed in logarithmic units relative to solar. For comparison, Table~\ref{tab:abundance_table} also includes an alternative estimate of [M/H] that is the average of the individual elemental abundances, which is not an uncommon calculation in the literature. Similarly, we compute the alpha-enhancement [$\alpha$/Fe] using the combined abundance of the $\alpha$-elements O, Mg, Si, Ca and Ti. We report two estimates: [$\alpha$/Fe] calculated using the \citet{Hinkel2022} formalism, and an arithmetic mean of [O/Fe], [Mg/Fe], [Si/Fe], [Ca/Fe] and [Ti/Fe].

\begin{figure*}
    \centering
     \includegraphics[width=\linewidth]{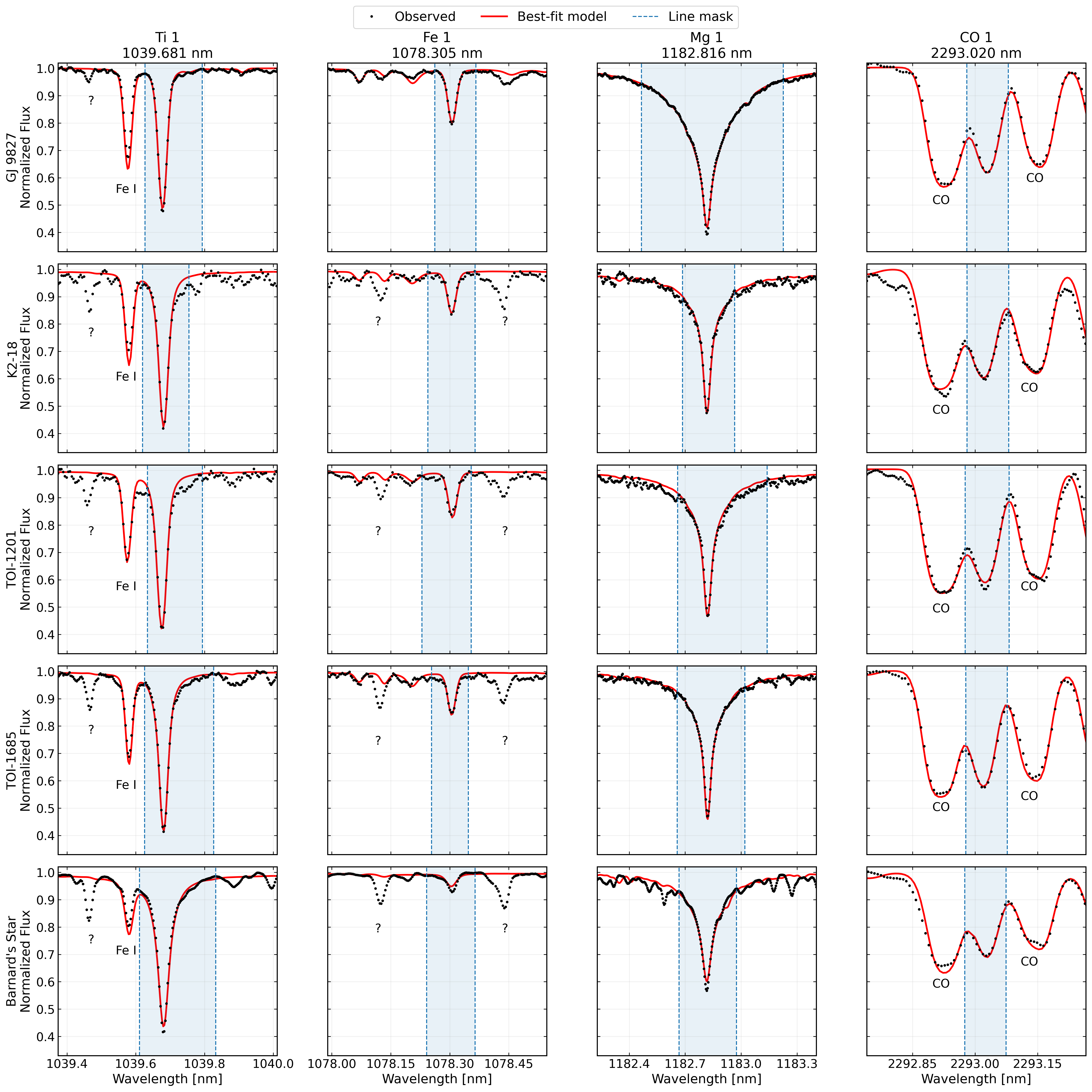}  
    \caption{Representative Ti I, Fe I, Mg I, and CO features for the five target stars. Observed spectra are shown as black points and best-fitting synthetic spectra as red curves. The blue shaded regions indicate the line masks used for fitting, while labels identify known or unknown nearby spectral features.}
    \label{fig:5x4plot}
\end{figure*}

Our bulk metallicities are broadly consistent with the literature values adopted in this work (Table~\ref{tab:stellarparams}). The one seemingly discrepant value is that of Barnard's Star, for which we derive [M/H] = $-0.48\pm0.15$~dex using a simple average over all [X/H] and $-0.33\pm 0.06$~dex using the prescription from \citet{Hinkel2022}. We note that the former is consistent with [M/H] reported by \citetalias{Jahandar2024}, who also equate their average [X/H] to [M/H]. Comparing to other studies of Barnard's Star's chemical abundances, reported [Fe/H] values span a wide range of -0.57 to -0.15 dex (see Section~\ref{sec:Barnard}), which we highlight is not equivalent to bulk metallicity. Figure~\ref{fig:Barnardplot} reveals that our measured value of [Fe/H] = $-0.722\pm 0.085$~dex is consistent with two out of four literature values. This level of disagreement is not alarming given that chemical abundance studies of Barnard's Star show large discrepancies for many elements and empirical calibrations are needed to assess the accuracy of the individual methods. Empirically calibrating our method presented herein using wide binaries is beyond the scope of this work and will follow in a forthcoming study. In Figure~\ref{fig:5x4plot}, we highlight representative fits for a selection of four spectral lines of Ti, Fe, Mg, and CO that are common to all five stars, illustrating the quality of our fits by comparing our best-fit models to the observed line. 

\input{Tables/abundancestable}

\section{Discussion} \label{sec:discussion}

\subsection{Elemental Trends} \label{sec:eletrends}
Our derived abundances and bulk metallicities are broadly consistent with literature measurements (see Sec. \ref{sec:litcomp}) and with the abundance patterns expected for nearby M dwarfs. We emphasize that the global metallicity calculated using the prescription from \cite{Hinkel2022} is not equivalent to [Fe/H], and can be strongly influenced by elements that dominate the total metal mass budget, such as oxygen. The observed spectral features used in this work are  well reproduced by the best-fitting synthetic spectra (c.f. Figure~\ref{fig:5x4plot}), indicating that our line selection and vetting processes successfully exclude many problematic blends and poorly modelled features. Nevertheless, some elements show notable deviations from expected chemical trends, most notably Na and K in TOI-1201 (c.f. Figure~\ref{fig:abun_plots-toi1201}) and Mn in K2-18 (c.f. Figure~\ref{fig:abun_plots-k2-18}). Being odd-Z elements, these elements are expected to exhibit decreased abundance trends with metallicity according to the Oddo-Harkins rule, which states that odd-Z elements are less abundant than even-Z elements due to nuclear stability and the prominence of alpha-capture reactions in stellar nucleosynthesis \citep{Kobayashi2006, Smiljanic_2016}. However Na, K, and Mn in the aforementioned stars exhibit the opposite trend of elevated abundances such that these outliers are more likely attributed to unidentified blends, inaccurate line transition data, or limitations in the atmospheric models. For example, certain elements have greater tendencies to depart from observations under the assumption of LTE whereby non-LTE corrections may be needed, in particular for K \citep{Olander2021}, Na \citep{Lind_2011}, and Ti \citep{Hauschildt_1997}.

As illustrated in Figure ~\ref{fig:errorplot}, the abundance uncertainties vary substantially among elements and targets. In many cases, the dominant contribution arises from systematic uncertainties rather than from line-to-line scatter, due to large errors on the adopted stellar parameters ($T_{\rm eff}$, $\log g$, [M/H], $v_{\rm mic}$, $v_{\rm mac}$). More precise characterization of stellar parameters therefore has the potential to result in substantial improvements in elemental abundance constraints. The availability and reliability of individual abundance diagnostics also vary strongly with \teff{}. In particular, commonly used Si and Ni features become progressively weaker or more blended with molecular absorption in cooler M dwarfs \citepalias[see Figure 16 in][]{Jahandar2024}. As such, fewer usable lines are available for these elements with decreasing \teff{,} and their abundances can deviate from expected trends as the individual features decrease in depth and the number of lines diminishes. This limitation becomes particularly important when deriving elemental ratios relevant to rocky planet mineralogy, such as Mg/Si \citep{Adibek_2015, BrewerFischerMgSiCO2016}. The large dispersion among published [Si/H] measurements for Barnard's Star (Figure~\ref{fig:Barnardplot}) illustrates the difficulty of obtaining robust Si abundances in the cool, metal-poor regime (i.e. $T_{\rm eff}=3231$~K, [M/H]$=-0.33$~dex). 

We additionally identify a systematic behaviour among some strongly broadened Mg, Al, and Ca features, which tend to yield lower abundances than narrower lines of the same species. This in turn can potentially produce bimodal distributions of [X/H] for the same element. We attribute much of this behavior to the increased difficulty of defining the local pseudo-continuum around broad or saturated features. Our PCN procedure (see Section~\ref{sec:PCN}) was designed and tested using both broad and narrow lines, and substantially reduces this discrepancy. Nevertheless, we acknowledge that broad and saturated features remain particularly sensitive to continuum placement \citep{Santos-Peral2020}. 

\begin{figure*}
    \centering    
    \includegraphics[width=\linewidth]{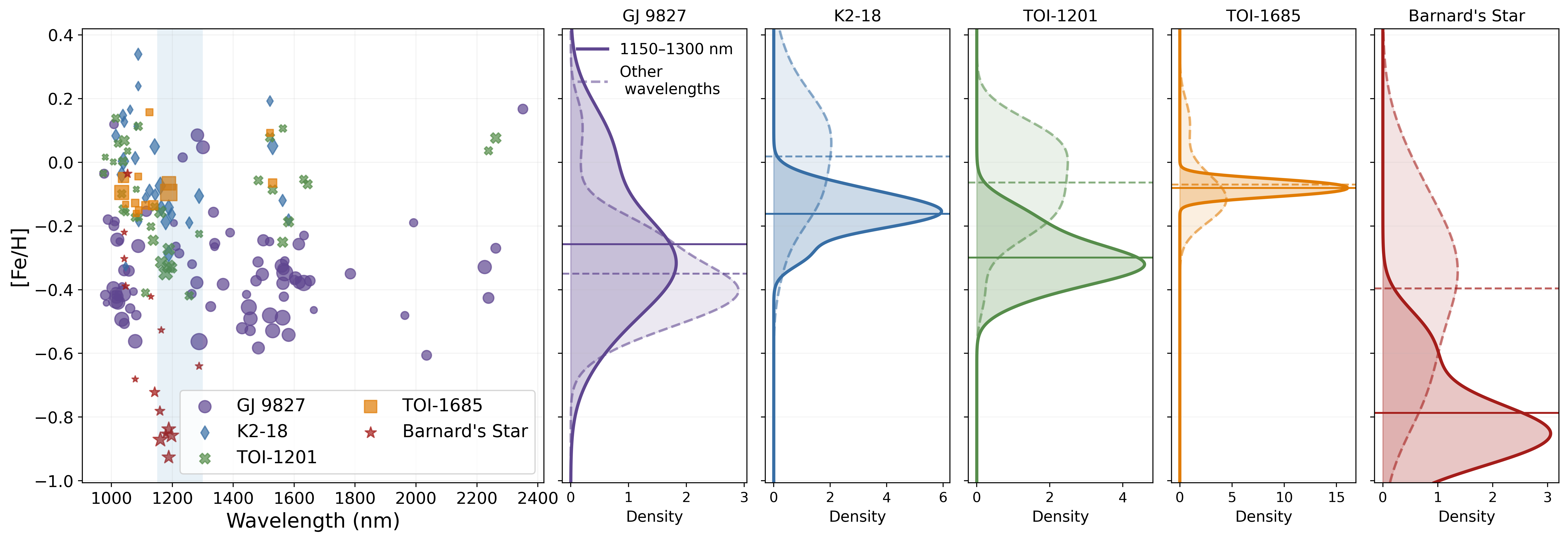} 
    \caption{\emph{Left}: Line-by-line [Fe/H] measurements as a function of wavelength for each star. The shaded region marks the 1150-1300~nm interval examined for the wavelength-dependent systematic. \emph{Right}: Kernel density estimates of the [Fe/H] distributions for the subsets of Fe I lines within 1150-1300 nm (solid) and at all other wavelengths (dashed) for each star. Horizontal lines indicate the corresponding mean abundances of each distribution.}
    \label{fig:Fe_wavelength_hists}
\end{figure*}

\begin{figure}
    \centering
    \includegraphics[width=\linewidth]{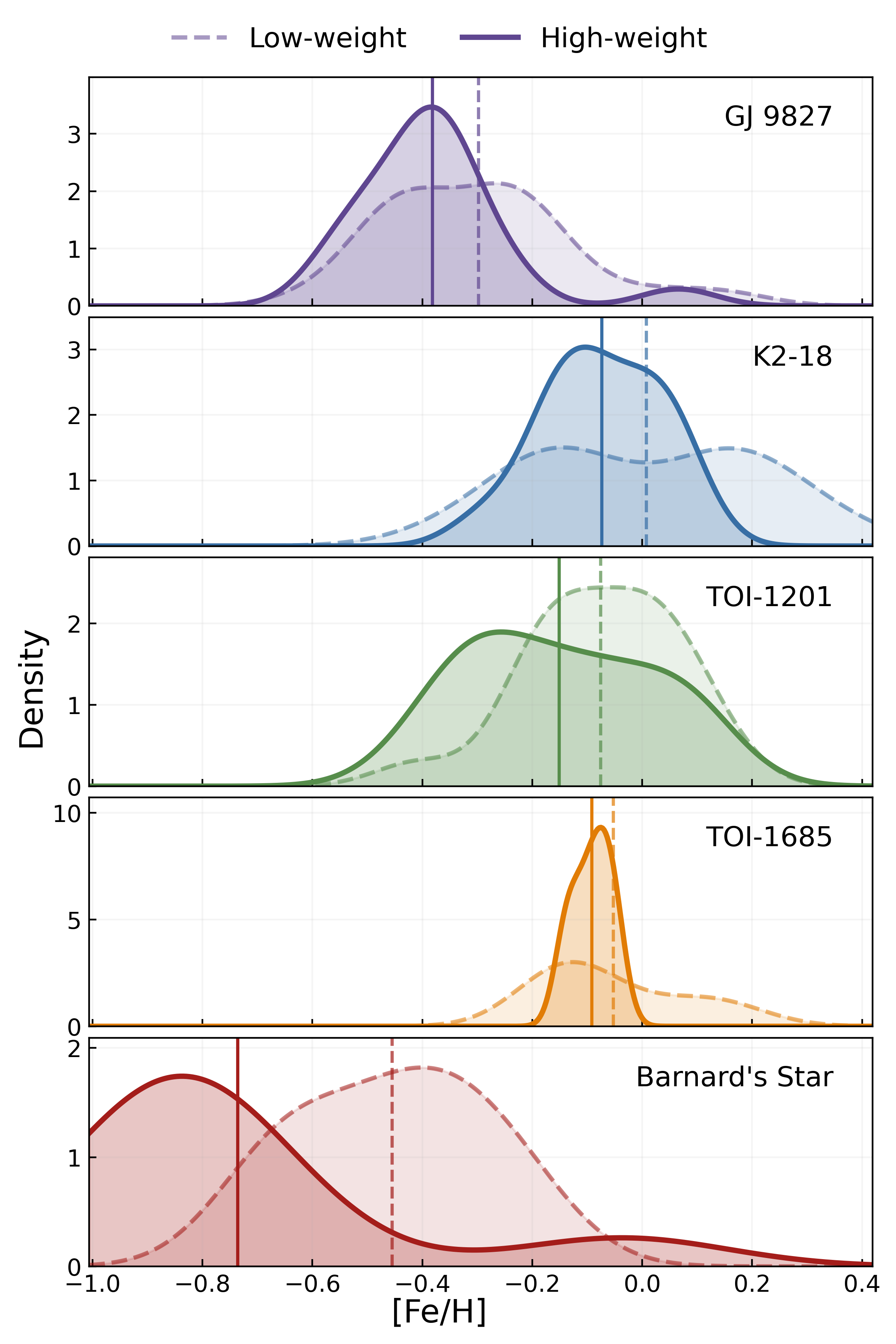} 
    \caption{Kernel density estimates of the [Fe/H] distributions for the high-weight (solid) and low-weight (dashed) halves of each star's Fe I linelist. Vertical lines indicate the corresponding mean abundances for each subset of lines.}
    \label{fig:Fe_weight_hists}
\end{figure}

A further systematic trend is seen among a subset of narrow Fe lines between approximately 1150 and 1300~nm. Lines within this wavelength range yield measurably lower [Fe/H] compared to other Fe lines for K2-18, TOI-1201, and Barnard's Star (Figure~\ref{fig:Fe_wavelength_hists}). This is despite all lines generally exhibiting high quality spectral fits and thus being assigned high weights when computing the weighted average [Fe/H]. To investigate whether this behavior is associated primarily with our adopted weighting scheme or whether it is truly a wavelength-dependent effect, we compare subsets of the per-line [Fe/H] distributions in Figures~\ref{fig:Fe_wavelength_hists} and~\ref{fig:Fe_weight_hists}. We first separate lines within 1150–1300 nm from all other lines and calculate the p-values of two-sample Kolmogorov–Smirnov (KS) tests. We find that spectral lines between 1150–1300 nm yield systematically lower [Fe/H] than lines elsewhere in the spectrum for K2-18 ($p=0.0092$), TOI-1201 (p=0.0003), and Barnard's Star ($p=0.013$) while no significant wavelength dependence is detected for GJ~9827, and the small number of Fe lines negates any meaningful assessment for TOI-1685. (c.f. Figure~\ref{fig:Fe_wavelength_hists}). Next, we bifurcate each star's [Fe/H] distribution into high-weight and low-weight subsets separated by the median weight. We find statistically significant differences in [Fe/H] between these subsets for GJ~9827 ($p=0.0013$) and Barnard's Star ($p=0.0025$), while the distributions for the remaining stars cannot be distinguished at the $p<0.05$ level. The differing results between the two tests suggest that the line-to-line [Fe/H] scatter cannot be attributed solely to our weighting prescription and that for K2-18, TOI-1201, and Barnard's Star, a wavelength-dependent systematic may contribute to lowering the recovered [Fe/H]. 

Importantly, this effect has little influence on our reported [Fe/H] values. We confirm this by replacing the weighted abundances with the median of all Fe lines and find that our resulting [Fe/H] values change by less than 0.05~dex for every target. The origin of any wavelength dependence remains unclear, but possible contributors include inaccurate oscillator strengths or damping constants, continuum placement, and deficiencies in the treatment of pressure broadening. In particular, recent work has shown that assumptions regarding van der Waals broadening can significantly affect synthetic spectra of cool M dwarfs, such that prescriptions calibrated primarily on Sun-like stars may introduce systematic errors at lower $T_{\rm eff}$ \citep{Glidden_2026}. We therefore interpret our KS tests as diagnostics of potential line- and wavelength-dependent modelling systematics present in the cooler stars of our sample, rather than evidence for intrinsically distinct abundance populations.

\subsubsection{Alpha Elements} \label{sec:alpha}
The alpha elements, produced during massive star evolution and in supernovae via $\alpha$-particle capture reactions, provide an important diagnostic of galactic chemical evolution \citep{Kobayashi2006, Kobayashi2020}. Within our sample we confirm the expectation that [$\alpha$/Fe] is generally enhanced towards lower [Fe/H], reflecting early alpha enrichment of the interstellar medium by core-collapse supernovae before delayed contributions of Fe emerge from Type Ia supernovae as the galaxy becomes increasingly populated with white dwarf Type Ia progenitors. Observations of the galactic disk consequently reveal that older, thick-disk stars are $\alpha$-enhanced relative to Fe, while thin-disk stars are only mildly $\alpha$-enhanced \citep{Bensby_2014, Hayden_2015, Weinberg_2019}. Because the targets in our sample belong to the thin disk, we expect solar or mildly enhanced alpha abundances \citep[\lbrack $\alpha$/Fe\rbrack $ \sim 0-0.2$~dex;][]{Bensby_2014}, with greater alpha enhancement plausible for increasingly metal-poor stars; particularly Barnard's star with [Fe/H]$=-0.722\pm0.085$~dex. 

Recall that the alpha elements in our study are defined as O, Mg, Si, Ca and Ti, as we ignore additional alpha elements whose abundances we do not measure (i.e. Ne, S, and Ar). These elements do not behave identically, however, with Ti in particular having a more complex nucleosynthetic origin than lighter alpha elements \citep{Kobayashi2006, Bensby_2014}. Our measured $[\alpha/\mathrm{Fe}]$ values in Table~\ref{tab:abundance_table} follow the expected trend. Barnard's Star, the most metal-poor star in our sample, shows the strongest evidence for enhanced $[\alpha/\mathrm{Fe}]=0.40\pm0.11$~dex, while the remaining stars exhibit $[\alpha/\textrm{Fe}]$ values consistent with solar at $<1.5\sigma$. As expected, the individual alpha element abundances are also reasonably consistent with [O/H], [Mg/H], and [Si/H] within $1\sigma$ for all five stars. Similarly, [Ca/H] is consistent for all stars except GJ~9827. Conversely, [Ti/H] exhibits the most complex behavior, particularly for TOI-1201 and Barnard's Star. Previous NIR studies of M dwarfs have shown large scatter and systematically lower values in [Ti/H] when compared to FGK stars \citep{Jonsson_2020, Souto_2022, Vilar_2025}. In the coolest M dwarfs (\teff{} $< 3400$ K), interpretation of Ti I lines is further complicated by molecular equilibrium: an increasing fraction of neutral Ti is incorporated into TiO, which depletes the availability of Ti I as an accurate diagnostic of [Ti/H] \citep{Ishikawa_2020}. This effect should not be interpreted as a depletion of the star's bulk Ti abundance, but rather as a partitioning of Ti between predominantly Ti I and TiO within the stellar photosphere. Departures from LTE may also introduce an extra source of uncertainty in Ti I line formation \citep{Hauschildt_1997}.

\begin{figure}
    \centering
    \includegraphics[width=\hsize]{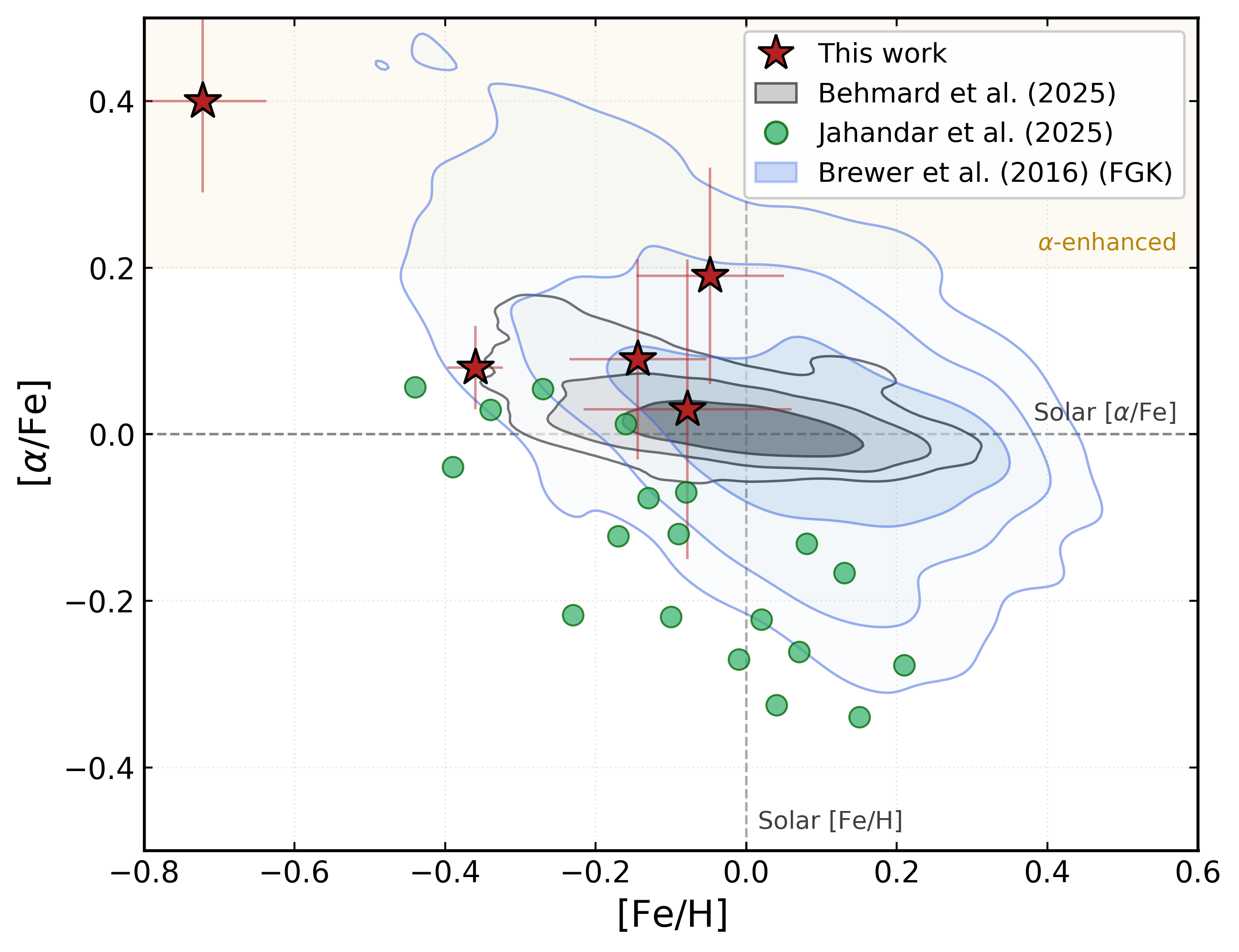}   
    \caption{[$\alpha$/Fe] versus [Fe/H] for the five stars in this work (red stars) compared to a sample of 31 nearby M dwarfs \citep{Jahandar_2025} and distributions of FGKM dwarfs from \citet{BrewerFischerCoolStars2016,Behmard_2025}. Dashed lines indicate the solar values and the shaded region marks strong alpha-enhancement with [$\alpha$/Fe].}
    \label{fig:alphaplot}
\end{figure}

Figure~\ref{fig:alphaplot} compares the [$\alpha$/Fe] values among our target stars with samples of FGKM stars  \citep{BrewerFischerCoolStars2016,Behmard_2025,Jahandar_2025}. The literature [$\alpha$/Fe] values depicted in Figure~\ref{fig:alphaplot} are not reported explicitly in these studies, so we calculate [$\alpha$/Fe] herein using the same definition of $\alpha$-enhancement as in our sample and adopting the same alpha-elements. Although systematic differences still remain between the underlying data and abundance measurement methodologies, we see that our stars occupy a similar region as the comparison samples. Barnard’s Star remains the notable exception given its elevated alpha-enhancement, driven by Fe being more depleted than the alpha elements, as recovered in our study as well as in \citet{Ishikawa2022,Olander_2025}.

\subsubsection{Iron-peak Elements} \label{sec:iron-peak}
Iron-peak elements are produced in both core-collapse and Type Ia supernovae, and are expected to broadly scale with [Fe/H] over the range of metallicities spanned by Galactic disk stars \citep{Kobayashi2006, Kobayashi2020}. Here we consider the iron-peak elements Cr, Mn, Sc, V, and Ni. Where available, per star [Cr/H] and [Ni/H] measurements are broadly consistent with each other and with [Fe/H], matching galactic trends previously recovered for both FGK and M dwarfs \citep{Souto_2022}. While our [Ni/H] measurements are limited to a small number of shallow Ni I lines, the broad wavelength coverage of SPIRou provides access to a relatively large number of clean Cr I lines (i.e. $\geq 10$). This makes Cr one of the best-constrained iron-peak elements in our analysis and the most reliable iron-peak indicator.

Sc and V exhibit more complex behaviour. Although they are considered iron-peak elements, their abundance trends have been reported to behave more similarly to alpha elements in stellar samples from the Galactic disk \citep{Batt&Bensby_2015}. Our [Sc/H] and [V/H] measurements appear to follow this behaviour, although we caution that the small number of lines for each element across our sample (i.e. $\leq 2$) makes drawing any firm conclusions difficult \citep{Adibek_2015}.

Mn is arguably the most difficult iron-peak element to interpret. One-dimensional LTE analyses of FGK dwarfs in the Galactic disc have recovered subsolar [Mn/Fe] at subsolar metallicities \citep[e.g.][]{Mishenina_2015, Ernandes_2018}, while non-LTE corrections have been shown to weaken or invalidate this trend in Sun-like stars \citep{Batt&Bensby_2015, Bergemann2019, Asplund_2021}. Although comparable calculations remain limited for M dwarfs, recent work suggests that non-LTE deviations are small for Mn I lines in APOGEE spectra of early-to-mid M dwarfs \citep{Souto_2022}. In addition, Mn and V are known to exhibit substantial hyperfine structure \citep{Shan_2021}. Some Mn and V lines are magnetically sensitive, such that Zeeman broadening in active M dwarfs will reshape these line profiles. If not accurately accounted for in atomic line data, the spectral synthesis models with the true abundances will appear too narrow and the derived abundances will consequently be biased to higher values. Although the stars in our sample are only mildly active, and the most magnetically sensitive of these lines were removed in the vetting process, we attribute this effect to the appearance of Mn enhancement in K2-18, TOI-1201 and TOI-1685, which we do not believe is real.

\subsection{Literature Comparison of Abundance Results}  \label{sec:litcomp}

Figures~\ref{fig:abun_plots-gj9827}, \ref{fig:abun_plots-k2-18}, \ref{fig:abun_plots-toi1685}, and \ref{fig:Barnardplot} compare our measured abundances to literature results where available. Our [X/H] measurements for TOI-1201 are the first such measurements to appear in the literature. For most of the remaining stars in our sample, our measured elemental abundances broadly match values recovered by previous studies, although some individual measurements show notable discrepancies. This is not necessarily unexpected, as these studies differ in wavelength coverage, spectral resolution, methodologies, model atmospheres, and line selection strategies \citep{Hinkel2016, BlancoC2019}. Here, we summarize these comparisons on a star-by-star basis and discuss likely sources of disagreement.

\subsubsection{Literature Comparison: GJ~9827}
One set of measured elemental abundances of GJ~9827 has been previously published by the JEWELS program, which focused on FGK stars \citep{Sun_2026}. JEWELS abundances were measured from 24 coadded optical HARPS \citep{Mayor_2003} spectra with an expected S/N = 270, using a line-by-line abundance analysis based on differential equivalent widths between GJ~9827 and the Sun \citep{Sun_2026}. Compared to our spectral synthesis-based method in the NIR, our derived abundances show large discrepancies from JEWELS (Figure~\ref{fig:abun_plots-gj9827}). This is despite JEWELS' relatively large uncertainties for most elements. These discrepancies may arise partly from the distinct wavelength regions and abundance diagnostics employed. Notably, \citet{Sun_2026} derives [C/H] from optical C$_2$ molecular bands, whereas our results rely on 176 CO lines, such that the inferred carbon abundances differ between these diagnostics \citep{Asplund_2021}. While our method has yet to be empirically calibrated, our results for nearly all elements are tightly clustered around the central [M/H] value and are more precise compared to the JEWELS results from optical spectroscopy, where many [X/H] values show large deviations from the central [M/H] and comparatively large uncertainties.

\subsubsection{Literature Comparison: K2-18}
A direct comparison against our measured abundances for K2-18 is provided by \citet{Hejazi_2024}. \citet{Hejazi_2024} present the \texttt{AutoSpecFit} framework, whose data and methodology closely match our own. Specifically, \texttt{AutoSpecFit} also performs a line-by-line spectral synthesis analysis using MARCS stellar atmosphere models and the Turbospectrum radiative transfer code. They perform an iterative $\chi^2$ squared minimization to derive [X/H] from high-resolution $HK$ spectra from IGRINS/Gemini-South. Given the similarities between our data and the \texttt{AutoSpecFit} methodologies, their [X/H] results for K2-18 provide the closest match to our results among any of the stars in our sample. As IGRINS similarly covers the $K$-band, they likewise use CO features to derive the carbon abundance. Despite these similarities, a few substantial differences exist between these two methodologies, such as the detailed treatment of local pseudo-continua and linelists. In particular, \citet{Hejazi_2024} use FeH features to derive [Fe/H], whereas we focus solely on Fe I lines. We do note that [Fe/H] inferred from Fe I versus FeH in cool star spectra have been shown to agree within $0.05$~dex in prior studies \citep{Souto2021}. Figure~\ref{fig:abun_plots-k2-18} compares our [X/H] values for K2-18 with those from \texttt{AutoSpecFit}. Our [Fe/H] value is consistent \texttt{AutoSpecFit} within $<1\sigma$ (i.e. $-0.05 \pm 0.10$ and $-0.03 \pm 0.11$, respectively). The difference in [Fe/H] values is 0.02~dex, indicating that the level of agreement in [Fe/H] is high between using Fe I versus FeH lines. Similarly, no other [X/H] value presented in both studies shows disagreement at a level $>1\sigma$. However, unlike in our study, \citet{Hejazi_2024} did not present measurements of [Si/H], [Mn/H], or [Cr/H], thus precluding a more thorough comparison. Fortunately, the missing elements are trace metals in terms of absolute abundances such that we can still compare the bulk metallicities between the two studies. We find a high level of agreement between our measured value of [M/H] and the value from \citet{Hejazi_2024} (i.e. $0.16\pm 0.07$ and $0.17\pm0.10$, respectively), although we achieve a slightly more precise measurement of [M/H] having accounted for the same error terms as \texttt{AutoSpecFit}.

\subsubsection{Literature Comparison: TOI-1201}
To our knowledge, no detailed elemental abundance analysis has yet to be published for TOI-1201. Our measurements herein provide the first extensive abundance measurements for this star and establish a reference to be used for future studies, including ongoing atmospheric observations with HST \citep{KreidbergHST}.

\subsubsection{Literature Comparison: TOI-1685}
Three independent studies have published measurements of TOI-1685's elemental abundances \citep{Hirano_2021,Melo2024,Behmard_2025}.
\citet{Hirano_2021} followed the methodology of \citet{Ishikawa_2020, Ishikawa2022}, which calculates line-by-line equivalent widths from Subaru/IRD spectra and spectral synthesis models. IRD features similar wavelength coverage and spectral resolution to SPIRou \citep[$YJH$ bands; $R\sim 70,000$;][]{Kotani2018}. As seen in Figure~\ref{fig:abun_plots-toi1685}, \citet{Hirano_2021} recover [X/H] for a limited set of elements (i.e. Na, Mg, Si, Ca, Ti, Cr, Mn, Fe), which generally show higher abundances and more scatter than our results, despite reported very similar levels of precision on the individual [X/H] values. While unconfirmed, these large discrepancies may reflect differences between our spectral synthesis-based versus the equivalent width-based analysis of \citet{Hirano_2021}. 

\citet{Melo2024} analyzed APOGEE spectra using spectral synthesis with MARCS stellar atmosphere models and Turbospectrum radiative transfer through the \texttt{BACCHUS} wrapper code. While APOGEE spectra are limited to the $H$-band and with lower spectral resolution than SPIRou (i.e. $R=22,500$ compared to $70,000$), the results from \citet{Melo2024} show good agreement with our results. This is likely, in part, due to our similar methodologies. Their uncertainty calculations also propagate the errors in the atmospheric parameters, as well as a 1\% and 2\% pseudo-continuum change. Notable differences include the use of both H$_2$O and OH lines to calculate [O/H], and both FeH and Fe I lines for [Fe/H]. These abundance measurements are consistent within $1\sigma$ between both studies. 

We also obtain good agreement with the data-driven abundances from \citet{Behmard_2025}. Their analysis uses the Cannon model, trained on APOGEE spectra of M dwarfs in wide binary pairs with FGK primaries of known compositions and applied to approximately 17,000 M dwarfs. Agreement between our findings and those from the calibrated, data-driven model of \citet{Behmard_2025} serves as an indicator that our TOI-1685 abundances are robust. The small uncertainties reported by \citet{Behmard_2025} primarily represent the empirical precision of their data-driven model, calibrated using repeat APOGEE observations, rather than a full propagation of stellar parameter and modelling systematics, as done here.

Comparing our [X/H] values with literature values is complicated by the wide range of metallicities reported for TOI-1685. Published estimates extend from [Fe/H] $\sim-0.20$~dex to $+0.30$~dex \citep{Bluhm_2021, Hirano_2021, Melo2024, Burt_2024, Gore_2024}, which encompasses our measured value of $-0.08\pm0.14$~dex. In particular, the adopted value of [Fe/H] = $0.14\pm 0.12$~dex reported by \citet{Hirano_2021} was obtained by combining several stellar-characterization methods. \citet{Melo2024} instead derive [Fe/H] $=0.06\pm 0.18$~dex, while \citet{Burt_2024} favours a substantially more metal-rich composition near $+0.3$~dex. This spread in [Fe/H], which propagates into input [M/H] in spectral synthesis models, illustrates how differences in underlying stellar parameters, and the assumptions that go into them, can impact the elemental abundances.

\subsubsection{Literature Comparison: Barnard's Star}
Barnard's Star is the most extensively studied object in our sample and is frequently used as a benchmark star for M dwarf abundance analyses. Values of [X/H] from these studies are shown in Figure~\ref{fig:Barnardplot}, each using a unique method. \citet{Maldonado2020} applied a model based on principal component analysis and sparse Bayesian regression to high-resolution optical spectra from HARPS and HARPS-N, using M dwarfs with FGK companions as a training sample. \citet{Ishikawa2022} measured equivalent widths from high-resolution NIR spectra from IRD and compared them to theoretical equivalent widths generated from synthetic spectra. Using the same SPIRou observations analyzed herein, \citetalias{Jahandar2024} fit a carefully vetted set of spectral features by interpolating PHOENIX-ACES stellar model grid over [M/H] as a proxy for [X/H] on a line-by-line basis. \citet{Olander_2025} used GIANO-B spectra \citep[$YJHK$, $R\sim 50,000$;][]{giano}, MARCS model atmospheres, and the \texttt{TSFitPy} line fitting package to perform a differential line-by-line analysis of Fe, Ti and Ca in M dwarfs, relative to the Sun. 

In comparison to the other, hotter stars in our sample, [X/H] values for Barnard's Star show significant variability across elements and between the aforementioned studies (c.f. Figure~\ref{fig:Barnardplot}). The RMS across the elements from individual studies is between $\approx 0.1-0.3$~dex, including our study at 0.14~dex. Similarly, individual [X/H] values can vary by as much as 0.6~dex, as is the case for [Si/H] between \citet{Maldonado2020} and \citetalias{Jahandar2024}, which reflects systematic differences between methods rather than line-by-line scatter alone. For the majority of measured elements, our derived abundances for Barnard's Star fall within the range of published results, although we do not consistently reproduce the abundances trends of any single study. Elements exhibiting the closest agreement with our results are generally obtained from either the same SPIRou data \citepalias{Jahandar2024} or from similar tools used \citep{Olander_2025}. The level of agreement remains element-dependent, suggesting that the dispersion in results cannot be solely attributed to differences in adopted stellar parameters. For example, our comparatively low [Fe/H] of $-0.72\pm0.09$~dex most closely matches those of \citet{Olander_2025} and \citet{Ishikawa2022}, but differs by $>3\sigma$ from those of \citetalias{Jahandar2024} and \citet{Maldonado2020}. Some elements, such as Mg, show relatively close inter-study agreement values whereas others, particularly Si, exhibit much greater dispersion. 
\citet{Maldonado2020} report systematically higher [X/H] for most species compared to the other four studies. Barnard's Star represents an especially challenging test case, as it is both the coolest and most metal-poor star in our sample. With \teff{}$=3231$~K, its NIR spectrum is dominated by molecular absorption from FeH, H$_2$O, TiO and other molecular opacity sources that can alter the local pseudo-continuum for atomic features. As a result, even with the exceptionally high S/N spectrum that we analyzed in this study, derived abundances remain susceptible to model-dependent errors that remain to be explored in future work.

While comparisons between independent abundances studies provide an important external check on the reliability of our results, they do not necessarily show that a given measurement is more or less accurate. High internal precision does not preclude stellar- and method-dependent systematic errors. Even with high S/N spectra at high resolution, differences in abundance measurements can disagree beyond their quoted uncertainties. The solar abundance scale illustrates the persistence of method-dependent systematics even in the highest-quality spectra: different analyses still recover measurably different solar abundances, despite the availability of meteoritic constraints from CI chondrites \citep{Lodders_2025}. Several sources of systematic error such as differences in adopted stellar parameters, atmosphere models, line selection, molecular opacity sources, the PCN, and departures from LTE, all contribute to the observed discrepancies, and these issues are generally magnified when applied to cooler stellar atmospheres \citep{Hinkel2016, BrewerFischerCoolStars2016, BlancoC2019, Jahandar2024}. In particular, non-LTE effects have been shown to be important for several species in abundance analyses \citep{Olander2021, LindAmarsi_2024, Olander_2025}, and remain insufficiently studied for many lines used in the NIR. Reported [X/H] values with uncertainties based only on random line-to-line scatter underestimate the total error budget when model and stellar parameter systematics have repeatedly been shown to be important \citep{Jofre2017, BlancoC2019}.
While our framework accounts for systematic uncertainties by perturbing input stellar parameters, our results remain model dependent. 

\subsection{Elemental Ratios}

Elemental abundance ratios encode information about the planet-forming chemical inventory that is not captured by overall metallicity or individual abundances. In particular, the relative abundances of C, O, Mg, Si, and Fe play key roles in setting the distributions of volatiles and major rock-forming species, atmospheric chemistry, the mineralogy of solid planet interiors, and the relative masses of silicate mantles to metallic cores, which influence the onset of plate tectonics and the potential for planetary habitability \citep{BrewerFischerMgSiCO2016, FoleyDriscoll_2016}. Elemental ratios therefore provide useful compositional priors for models of planetary interiors and atmospheres. Because stars and their protoplanetary disks form from the same natal material, stellar photospheric abundances are often used as proxies for the initial bulk composition of the planet-forming disk \citep{Bond_2010, Thiabaud2015, Dorn_2017, Adibek_2024}. 
Nevertheless, stellar and planetary compositions are not expected to correspond one-to-one as disk chemistry, condensation, radial drift, gas and solid accretion, differentiation, and subsequent planetary evolution can all modify a planet's final composition \citep{Oberg2023}. As such, the stellar elemental ratios presented herein are better interpreted as the initial chemical environment from which planets formed, rather than as direct measurements of the present-day planets. Table~\ref{tab:ratio_table} lists the C/O, Fe/Mg, Mg/Si and Fe/O molar ratios derived for our stellar sample. These values are compared with the corresponding solar ratios \citep{Asplund2009} and measurements for nearby samples of M and FGK dwarfs \citep{BrewerFischerMgSiCO2016, Jahandar_2025}.

\input{Tables/abundanceratios}

\subsubsection{Refractory Ratios} \label{sec:refractories}

Alongside O, the refractory elements Mg, Si, and Fe constitute most of the bulk material in rocky planetary interiors. Mg and Si are primarily incorporated into the silicate mantle with Fe being heavily partitioned into a non-pure metallic core, with some Fe also present in the mantle. Host star Fe/Mg ratios serve as a constraint on the Fe inventory available to form planetary cores relative to mantle-forming materials, and have been used as priors on the iron core mass fractions (CMF) of super-Earths and sub-Neptunes \citep{Dorn_2015, Cadieux_2022, Cadieux_2024}. We note that Fe/Mg is favored over Fe/Si as a proxy for the core-to-mantle ratio in solid planetary interiors given that Si is more siderophile than Mg at typical core:mantle conditions. This gives Mg less affinity to partition into a metallic core than Si, and therefore Mg remains highly representative of a planet's mantle portion. Similarly, host star Mg/Si ratios contribute to governing the mantle mineralogies of small planets with Mg/Si $>1$ producing free MgO whereas Mg/Si $<1$ produces free SiO$_2$. These compositional differences modify the mantle's equation of state and set its internal density profile. Because these elements condense at relatively high temperatures \citep[$T_c > 1310$ K;][]{Lodders_2003}, they solidify early in the history of the disk, making them available to be accreted into solid protoplanetary cores. Furthermore, because these elements have very high condensation temperatures, planet formation models generally predict that these refractory ratios are preserved between a host star and its planets, making them a powerful diagnostic tool of planetary composition \citep{Bond_2010, Thiabaud2015, Dorn_2017, HinkelUnterborn_2018, Jorge_2022, Sanchez_2026}. 

\begin{figure}
    \centering
    \includegraphics[width=\hsize]{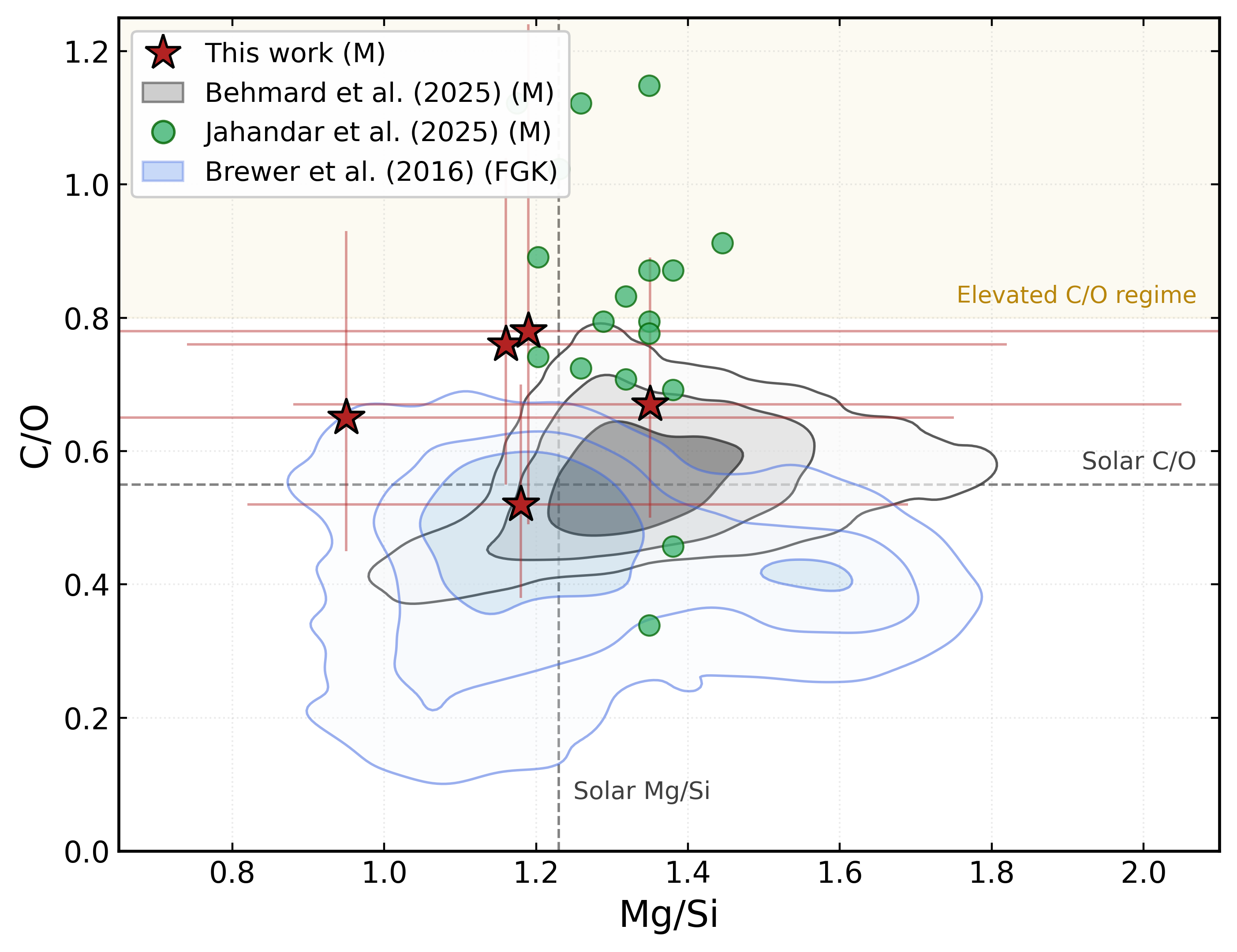}
    \caption{C/O as a function of Mg/Si for the five stars in this work (red stars) compared with \citet[][green circles]{Jahandar_2025} and the distributions calculated using values from \citet[][grey contours]{Behmard_2025} and \citet[][blue contours]{BrewerFischerCoolStars2016}. Dashed lines indicate the adopted solar C/O and Mg/Si ratios, and the shaded region marks C/O > 0.8 \citep{Bond_2010}.}
    \label{fig:COMgSiplot}
\end{figure}

Our recovered Mg/Si ratios range from 0.95-1.35 and are all consistent with the solar value within $1\sigma$. This is largely driven by the large uncertainties in [Mg/H] and [Si/H] for cool stars due to the small number of reliable Mg and Si lines and their strong temperature dependence (c.f. Figure~\ref{fig:errorplot}), which is dictated by input \teff{} and its measurement uncertainty. The Mg/Si ratios for all five stars in our sample are also consistent with the Mg/Si distributions for M dwarfs \citep{Behmard_2025,Jahandar_2025} and FGK stars \citep{BrewerFischerMgSiCO2016} as shown in Figure~\ref{fig:COMgSiplot}. All of our targets have Mg/Si ratios that are consistent within $1\sigma$ with the intermediate $1<\mathrm{Mg/Si}<2$ regime. Here, planetary mantles contain Mg that is expected to be evenly mixed between forsterite (Mg$_2$SiO$_4$) and enstatite (MgSiO$_3$) phases in simple stoichiometric models \citep{CarterBond_2012, BrewerFischerMgSiCO2016}. We also find no strong evidence for the Mg-rich compositions associated with Mg/Si $>2$. As shown in Table~\ref{tab:Error_table} and Figure~\ref{fig:errorplot}, the Mg and Si abundances are highly sensitive to \teff{,} particularly in the coolest stars, which drives up uncertainties on our measured Mg/Si ratio such that Mg/Si $<1$ is not strongly ruled out for any of our targets. 

\begin{figure}
    \centering
    \includegraphics[width=\hsize]{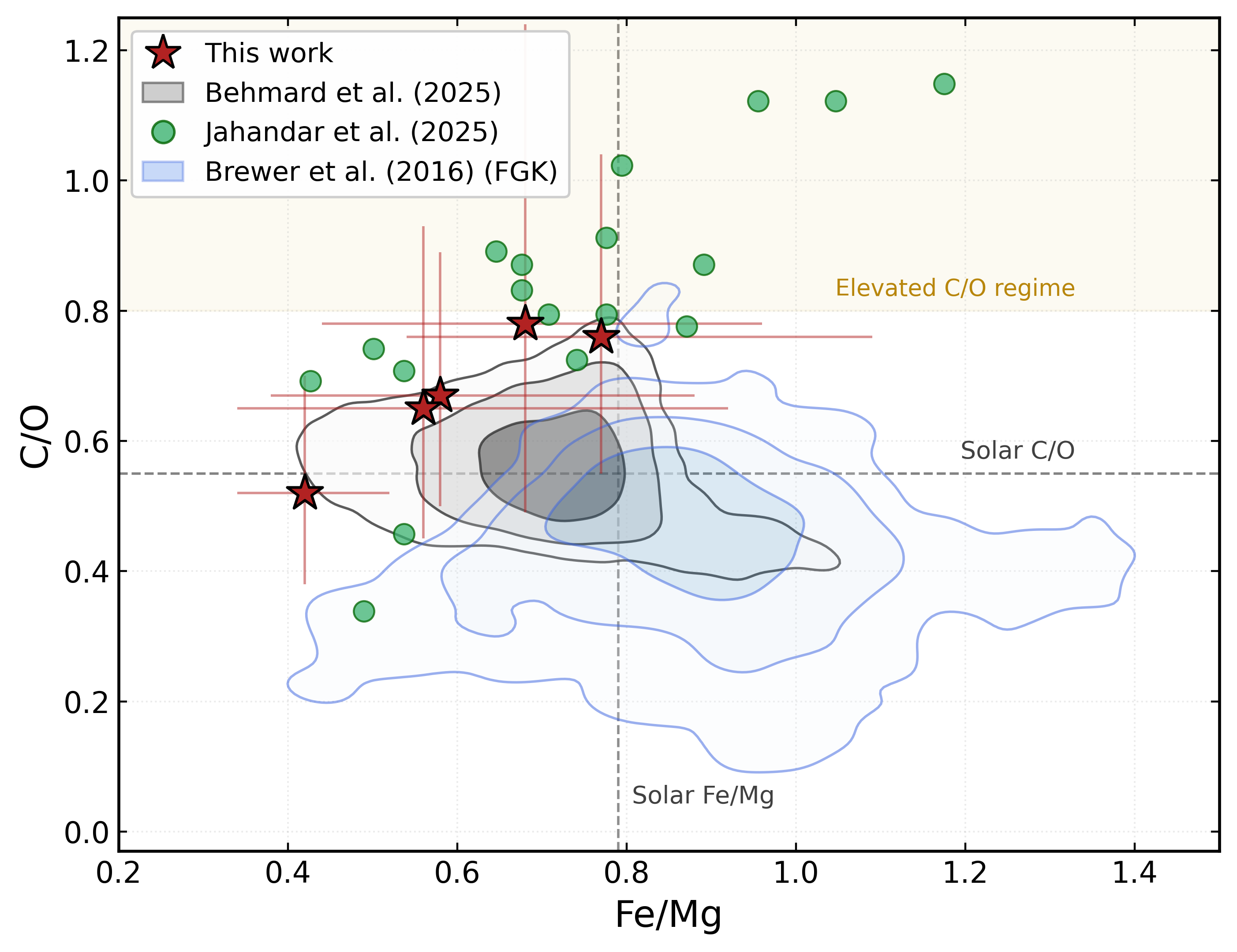}   
    \caption{Same as Figure~\ref{fig:COMgSiplot}, but for C/O versus Fe/Mg.}
    \label{fig:COFeMgplot}
\end{figure}

The Fe/Mg ratios of all five targets fall below the solar value of $0.79$ with varying degrees of significance from $0.1-2.2\sigma$, with every star other than Barnard's Star being consistent with solar within $1\sigma$. Barnard's Star exhibits the lowest Fe/Mg=$0.42^{+0.10}_{-0.08}$ as a result of its low [Fe/H]$=-0.722\pm 0.085$. 
Though Figure~\ref{fig:COFeMgplot} shows that  Barnard's Star's Fe/Mg ratio remains broadly consistent with the Fe/Mg distributions of M dwarfs \citep{Behmard_2025,Jahandar_2025}.  
While physical effects in planetary interiors such as radial chemical transport, giant impacts, differentiation, and volatile delivery can all modify the idealized 1-to-1 connection between host star and planetary compositions, 
a low stellar Fe/Mg ratio would favor a smaller planetary CMF. Core size and composition are vital when considering conditions for habitability through dynamo generation, the strength of the resulting magnetic field, and the onset of tectonic activity \citep{FoleyDriscoll_2016}. 

Similarly, the Fe/O ratio can help constrain the availability of oxygen to incorporate iron into oxidized mantle phases \citep{Jorge_2022}. The Fe/O ratios of our targets range from 0.03 to 0.06 and are consistent within $1\sigma$ of the solar value ($0.07\pm0.02$), with the exception of Barnard's Star (i.e. Fe/O= $0.03\pm0.01$). This feature of Barnard's Star is the result of its depleted [Fe/H] relative to all other elements. The measurements of Fe/Mg and Fe/O should be interpreted in light of the unusual Fe abundance behaviour identified in Section~\ref{sec:eletrends}, as a systematic underestimate of [Fe/H] would bias both inferred Fe/Mg and Fe/O ratios to be lower.

\subsubsection{Volatile Ratios}  \label{sec:volatile}

C and O are major constituents of the volatile material in protoplanetary disks. The differential freeze-out of molecular species such as H$_2$O, CO$_2$, CO, and others across their respective snow lines produce strong radial variations in disks' gas- and solid-phase C/O ratios \citep{Oberg2011, Eistrup_2016}. Atmospheric C/O for gas-rich planets has consequently been proposed as a potential tracer of formation location and migration \citep{Pontoppidan_2014}. However, the stellar and planetary C/O ratios are not expected to directly correspond to one another given that time-dependent disk chemistry, condensation, disk age, vertical mixing in the planetary atmosphere, and the relative accretion of gases, ices, and refractory solids during migration determine what material forms present-day planetary atmospheres \citep{Thiabaud2015, Eistrup_2016, Molliere_2022, Bergin_2024, Souto_2026}. Furthermore, much of the classical interpretations of atmospheric C/O relative to the host star were developed for giant planets \citep{Konopacky_2013, Reggiani_2022, Drazkowska2023}, but the aforementioned distinctions are especially relevant for sub-Neptunes, whose atmospheric C/O may be altered by chemical exchange with a molten interior or a magma ocean \citep{SeoIto_2024, Werlen_2025}. Planet host star C/O ratios therefore provide estimates of the system's initial bulk inventory of C and O, which helps contextualize the interpretation of planetary C/O ratios rather than providing a direct probe of formation pathways. Such connections may be strengthened with additional volatile abundances, for instance, C/N is believed to provide useful constraints when combined with C/O as major N-bearing species condense at temperatures distinct from those of the major C- and O- carriers, thus distinguishing between gas and solid accretion in different disk regions \citep{Cridland_2020, Turrini2021, Pacetti_2022}. Sulphur can also act as a complementary tracer \citep{Crossfield_2023}. However, we are unable to report [N/H] or [S/H] in this work as the available S I, N I, and CN lines are too weak and heavily blended in our targets' spectra and are not adequately reproduced in our synthetic spectra at this time. 

Stellar C/O may also influence the mineralogy of rocky planet-building material. Early studies adopted C/O $\simeq0.8$ as a threshold, above which there is not enough O available to form the silicate-dominated compositions seen in the solar system \citep{KuchnerSeager_2005, Bond_2010}. Instead, planetary interiors heavily favored carbide- and graphite-rich solids \citep{KuchnerSeager_2005, Bond_2010}. Recent condensation calculations indicate a more gradual transition in which silicate abundances begin to decline for C/O $\gtrsim0.65$, and strongly carbide-dominated systems emerge closer to C/O $\sim0.9$ \citep{Moriarty_2014, Shakespeare_2025}. Additionally, at C/O values $>0.7$, Mg/Si and Fe/Mg are expected to deviate substantially from the host stellar value, as the increasing C/O lowers the oxygen fugacity and changes the condensation sequence of Si, Mg and Fe \citep{Zaveri_2026}. However, there is still heavy speculation in how common these carbon-rich systems are in nature. While early spectroscopic surveys reported having high C/O $>0.8$ in approximately 25\%–30\% of systems, and C/O $>1.0$ in 6\%–10\% \citep{Delgado2010, PetiguraMarcy_2011}, more recent studies have demonstrated that FGK star C/O ratios are more commonly consistent with the solar value than was previously reported \citep{Fortney_2012, Nissen_2013, Nissen_2014, BrewerFischerMgSiCO2016, Bedell_2018, Stonkute_2020}.

Our measured stellar C/O ratios range from 0.52 to 0.78 and are all consistent with the solar value within $1\sigma$. We note that the C/O ratios for GJ~9827, K2-18, TOI-1201, and TOI-1685 are systematically higher than solar, and are also
moderately higher than the sample mean of FGK stars from \citet{BrewerFischerMgSiCO2016}, while being compatible with the larger mean C/O reported for the M dwarf sample of \citet{Jahandar_2025}, which results from a significant subsample of their M dwarf sample being purported to have elevated C/O $\geq0.8$ (c.f. Figures~\ref{fig:COMgSiplot} and \ref{fig:COFeMgplot}). None of our stars show evidence for carbon-rich chemistry needed to produce carbide-rich planets \citep[C/O $\geq 0.8$;][]{KuchnerSeager_2005}. 
As shown in Figure~\ref{fig:COFeMgplot}, our targets' C/O ratios lie near the upper end of the comparison distributions for FGK and M dwarfs, rather than being clear outliers. Our moderately elevated C/O values could partially reflect a systematic behavior when deriving [C/H] from CO lines rather than being indicative of a true chemical enrichment. The OH features used to derive [O/H] are generally well reproduced by the models and the resulting [O/H] values follow broad alpha-element patterns, as discussed in Section~\ref{sec:alpha}. On the other hand, because some CO lines are saturated and sensitive to \teff{}, $v_{\rm mic}$, and $v_{\rm mac}$, [C/H] does show greater line-to-line dispersion than [O/H] from OH lines and no reliable C~I features are available in our spectra to provide an independent, atomic diagnostic for [C/H] \citep{Asplund_2021}.

\subsection{Planet Formation Implications} \label{sec:planets}

\subsubsection{Rocky Planet Interiors} \label{sec:rockyplanets}
Stellar refractory ratios are most directly relevant to the rocky sub- to super-Earths in our sample: GJ~9827 b, c, TOI-1685 b, and the four non-transiting sub-Earth planets around Barnard's Star \citep{Basant_2025}. Because the planets of Barnard's Star do not transit it, their radii and bulk densities are not known. While their low masses and high irradiation suggest that they are small, rocky planets, it is the stellar composition that provides one of the few possible constraints on their interiors. \citet{Byrne_2026} analyzed the interiors of the sub-Earths orbiting Barnard's Star conditioned on its purportedly super-solar Mg/Si ratio \citepalias[Mg/Si$=2.63\pm 0.52$;][]{Jahandar2024} and predicted ferropericlase-rich mantles with reduced water-storage capabilities. Our analysis reveals a significantly lower Mg/Si$=1.18^{+0.51}_{-0.36}$, consistent with solar, which does not support the Mg-rich stellar prior adopted for this mineralogical prediction, and would therefore revise the inferred mantle composition. Our low Fe/Mg also nominally favors relatively small iron CMFs in these planets, but such a conclusion is not testable given their unknown planetary radii. While sub-Neptunes have been shown to preferentially form around M dwarfs with high [Fe/H] compared to terrestrial planets \citep{Turtelboom_2026}, Barnard's Star's low Fe/Mg likely does not preclude the formation of terrestrial bodies as alpha-elements are known to partially compensate for a reduced availability of Fe in terrestrial planet formation \citep{Adibek2012b, Steffen_2025}.

GJ~9827 b and c have masses and radii consistent with approximately Earth-like compositions \citep{Passegger2024}. The star's solar-like Mg/Si ratio of $1.35^{+0.70}_{-0.47}$ is compatible with the interpretation of an Earth-like mixture of silicate phases for planets b and c.

Despite early classifications of TOI-1685 b as a volatile-rich world, RV follow-up efforts revealed the planet has an approximately Earth-like composition \citep{Burt_2024}. JWST phase-curves and transmission observations of TOI-1685~b favor a dark, barren surface and rule out a clear H$_2$-dominated atmosphere, likely due to XUV-driven atmospheric escape \citep{Luque_2024, Fisher_2026}. However, a heavy secondary atmosphere remains consistent with the atmospheric observations. Assuming that TOI-1685 b lacks any substantial surface layer of volatiles, we calculate its iron CMF using \texttt{exopie} \citep{Plotnykov2024} based on the planet's mass and radius from \citet{Burt_2024}. We find that CMF$_p=0.12_{-0.08}^{+0.13}$. Compared to the planetary CMF predicted by our measurements of [Mg/H], [Si/H], and [Fe/H], we find a moderately larger value of CMF$_\star=0.24_{-0.11}^{+0.12}$. While the planet's expected CMF$_\star$ exceeding CMF$_p$ is not statistically significant for this one system, this finding is consistent with the population-level trend that hot super-Earths around M dwarfs have systematically smaller CMF$_p$ values than predicted by their host stars; a trend that has been interpreted as evidence for sequestered water with mass fractions of $\sim1$\% \citep{Weisserman_2026}. 

\subsubsection{Volatile-Rich Planets} \label{sec:volatileplanets}
The stellar C/O ratios are most directly useful as reference values for the volatile-rich planets in our sample: GJ~9827~d, K2-18~b, and TOI-1201~b, which are targets for atmospheric characterization with HST and JWST. JWST/NIRISS observations have revealed that GJ~9827~d possesses a highly metal-enriched, steam-dominated atmosphere \citep{PiauletGhorayeb_2024}. This result establishes the presence of a large, O-bearing volatile reservoir that is depleted in C. While potentially due to geochemical interactions between the planet's atmosphere and interior, our measured host star C/O argues against the host system being intrinsically carbon-poor, implying that the planet's C depletion instead reflects disk-scale fractionation and/or subsequent atmospheric-interior processing (Piaulet-Ghorayeb et al. in prep.).

K2-18 b currently provides a more complicated comparison. While recent analyses with JWST robustly recover CH$_4$ in its atmosphere \citep{Madhu2023_DMS}, they disagree on the presence of oxygen-bearing species and on the recovered atmospheric C/O, depending on which data reductions and retrieval models are used. For example, \citet{SchmidtMacDonald_2025} and \citet{Jaziri_2025} find no support for CO$_2$, favoring the picture of an O-poor sub-Neptune with high metallicity and super-stellar C/O. Our stellar value of $0.65_{-0.20}^{+0.28}$ serves as a reference for these retrievals, but the present atmospheric data do not yet firmly establish whether the planet is genuinely super-stellar in C/O. If confirmed, any differences would provide evidence for selective gas or solid accretion, or interior--atmosphere processing, and help constrain potential formation pathways.

While the mass and radius of TOI-1201~b imply a significant volatile component, these data alone are insufficient to distinguish between a water-rich interior and a H/He-enveloped gas dwarf \citep{Kossakowski_2021}. No atmospheric measurements have yet been published, such that the planet's atmospheric C/O and metallicity remain unknown. The slightly super-solar C/O$=0.76^{+0.46}_{-0.29}$ reported herein represents a useful point of comparison for future atmospheric characterization.

\section{Summary \& Conclusions} \label{sec:summary}
We present our framework for a line-by-line elemental abundance analysis of cool stars based on fitting spectral synthesis models to high-resolution, near-infrared spectra. We apply our technique to SPIRou spectra of five planet-hosting cool stars, GJ~9827, K2-18, TOI-1201, TOI-1685, and Barnard's Star, and derive abundances of $\geq 12$ elements for each star. Our main conclusions are summarized below.

\begin{itemize}
    \item We outline our methodology to measure line-by-line elemental abundances of cool stars. The steps include wavelength alignment, global continuum and local pseudo-continuum normalizations, line selection--including atomic lines and the use of CO and OH lines to measure [C/H] and [O/H], respectively--our treatment of nuisance line broadening parameters, $\chi^2$-minimization to measure [X/H] using spectral synthesis models, and a detailed analysis of random and systematic errors.
    \item We measure the abundances of C, O, Na, Mg, Al, Si, K, Ca, Ti, Cr, Mn, and Fe for all five of our target stars, plus Sc, V, and Ni for different subsets of stars. From the individual [X/H] measurements we recover each star's overall metallicity [M/H] and [$\alpha$/Fe].
    \item We compare our results against literature values, where available.
    \begin{itemize}
        \item For GJ 9827 ($T_{\rm eff}=4236$~K), we derive precise constraints ($\lesssim 0.1$ dex) on most [X/H] values, [M/H], and [$\alpha$/Fe], with [M/H] being consistent with literature values.
        \item For K2-18 ($T_{\rm eff}=3547$~K), we find close agreement with the results from \citet{Hejazi_2024} for nearly all elements. \citet{Hejazi_2024} represents the closest analog to our work in terms of data used and methodology.
        \item For TOI-1201 ($T_{\rm eff}=3659$~K), our measured abundances are the first such measurements in the literature.
        \item For TOI-1685 ($T_{\rm eff}=3519$~K), we find strong agreement for all [X/H]--except Mn--between the spectroscopic studies of \citet{Melo2024} and \citet{Behmard_2025}. In contrast, the [X/H] results from \citet{Hirano_2021} exhibit systematically higher [X/H] compared to ours and the aforementioned studies.
        \item For Barnard's Star ($T_{\rm eff}=3231$~K), the coolest and most metal-poor star in our sample, we compare our [X/H] values against four similar studies, which show large inter-study variations in [X/H] and large intra-study variations across the elements. This highlights the difficulty in measuring accurate abundances from high-resolution spectroscopy of cool, metal-poor stars, even at extremely high S/N$\gtrsim 1000$. Our [X/H] results generally sit between literature values.
    \end{itemize}
    \item We also calculate the elemental ratios C/O, Fe/Mg, Mg/Si, and Fe/O, which have implications for the atmospheres and solid interiors of orbiting planets. We find that all of these ratios are consistent with the solar values within $1\sigma$, with the exception of Fe/Mg and Fe/O for Barnard's Star, both of which are subsolar due to the star's low [Fe/H]$=-0.72\pm0.09$~dex. 
    \item We refine the Mg/Si ratio of Barnard's Star, which was previously reported to be $2.2\times$ larger than our reported value of Mg/Si$=1.18^{+0.51}_{-0.36}$. This finding significantly modifies past mineralogical predictions for the compositions of the four known sub-Earths in the system.
\end{itemize}

The spectral synthesis techniques applied to high-resolution spectra provide a powerful toolkit to assess star-planet compositional connections. Instruments like SPIRou can serve as workhorses for the analysis of cool star elemental abundances from NIR spectra. Due to the small size and heterogeneous nature of our stellar sample, we do not attempt to make or confirm any population-level conclusions regarding host star properties or trends with planetary properties. Instead, our work establishes detailed stellar abundance baselines for benchmark planetary systems around cool stars and demonstrates a methodology for deriving precise elemental abundances from NIR spectroscopy. Future work will empirically calibrate the results from our framework, enabling direct assessment of method-dependent abundance offsets and the external accuracy of the inferred abundances. Applying our analysis to a larger sample, alongside improved atomic and molecular data and detailed measurements of planetary atmospheres and interiors, will provide stronger tests of how closely planets inherit the chemical compositions of their host stars.

\begin{acknowledgments}

Acknowledgements: We thank Emma Turtelboom for helpful discussions. This research was enabled in part by support provided by BC DRI Group and the Digital Research Alliance of Canada (alliancecan.ca). DW acknowledges the support of the Natural Sciences and Engineering Research Council of Canada (NSERC). RC acknowledges the support of the Natural Sciences and Engineering Research Council of Canada (NSERC) and the Canada Research Chairs (CRC) Program [Award ID: CRC-2024-00372].  AL acknowledges support from the Fonds de recherche du Québec -- Secteur Nature et technologies (FRQNT) under grant \#349961. CC acknowledges the support from the Swiss National Science Foundation under the grant SPECTRE (No. 200021\_215200).

We recognize that McMaster University is located on the traditional territories of the Mississauga and Haudenosaunee nations, and within the lands protected by the “Dish with One Spoon” wampum agreement. This work is based on observations obtained at the Canada-France-Hawaii Telescope (CFHT), located on Maunakea in Hawaii. We acknowledge the considerable cultural, natural, and ecological significance of Maunakea, a sacred site to Native Hawaiians, and recognize the privilege of conducting astronomical research using observations obtained from this mountain.
\end{acknowledgments}

%

\vspace{5mm}
\facilities{CFHT/SPIRou}


\software{
iSpec \citep{BlancoC2014, BlancoC2019},
Turbospectrum \citep{Alvarez1998, Plez2012},
MARCS \citep{Gustafsson2008},
VALD \citep{Ryabchikova_2015},
APERO \citep{Cook2022},
exopie \citep{Plotnykov2024},
NumPy \citep{Harris_2020},
SciPy \citep{Virtanen2020},
Matplotlib \citep{Hunter_2007},
Astropy \citep{Astropy2013, Astropy2018, Astropy2022}
}


\clearpage
\appendix

\renewcommand{\thetable}{\Alph{section}.\arabic{table}}
\counterwithin{table}{section}

\section{Complete Record of Abundances with Errors}
Here we present the complete results of the abundance analysis for each of the five target stars, including the individual contributions of each error term and the number of lines used for each star.

\input{Tables/abundanceerrors}
\FloatBarrier

\input{Tables/abundanceerrors_cont}
\FloatBarrier

\counterwithin{table}{section}

\section{Complete Linelists for Target Stars} \label{sec:full_linelist}
This appendix lists the spectral features used in the abundance analysis, grouped by species and identified by their central wavelengths. For each target, the corresponding entry gives the abundance inferred from that line when the feature was retained in the final linelist. The complete table is included in the arXiv source files and will be included as a machine-readable table in the published version.
\FloatBarrier

\input{Tables/linelist_shortened}

\bibliography{citations}{}
\bibliographystyle{aasjournal}

\end{document}

%% file: Tables/stellarparams.tex
\begin{table*}[t]
\caption{Stellar parameters for targets}
\centering
\label{tab:stellarparams}

\footnotesize
\setlength{\tabcolsep}{0pt}
\renewcommand{\arraystretch}{0.75}

\begin{threeparttable}
\begin{tabular*}{\textwidth}{@{\extracolsep{\fill}}lccccc}
\toprule
Parameter & GJ 9827 & K2-18 & TOI-1201 & TOI-1685 & Barnard's Star \\
\midrule


TIC ID & 301289516 & 388804061 & 29960110 & 28900646 & 325554331 \\
2MASS ID & J23270480--0117108 & J11301450+0735180 & J02485926--1432152 & J04342248+4302148 & J17574849+0441405 \\

R.A. (J2000)\tnote{a} &
23:27:04.84 &
11:30:14.52 &
02:48:59.27 &
04:34:22.50 &
17:57:48.50 \\

Dec (J2000)\tnote{a} &
-01:17:10.58 &
+07:35:18.26 &
-14:32:14.94 &
+43:02:14.69 &
+04:41:36.11 \\

Distance (pc)\tnote{a} &
$29.6519 \pm 0.0148$ &
$38.100 \pm 0.0386$ &
$37.6356 \pm 0.0317$ &
$37.609 \pm 0.0272$ &
$1.8282 \pm 0.0001$ \\

Spectral type &
K7V\tnote{b} &
M2.5V\tnote{g} &
M2V\tnote{l} &
M2V\tnote{l} &
M4.2V\tnote{o} \\

$T_{\mathrm{eff}}$ (K) &
$4236 \pm 12$\tnote{c} &
$3547 \pm 85$\tnote{h} &
$3659 \pm 73$\tnote{l} &
$3519 \pm 100$\tnote{n} &
$3231 \pm 21$\tnote{p} \\

$[\mathrm{M/H}]$ (dex) &
N/A &
N/A &
$-0.121 \pm 0.082$\tnote{l} &
$0.088 \pm 0.082$\tnote{l} &
$-0.48 \pm 0.04$\tnote{p} \\

$[\mathrm{Fe/H}]$ (dex) &
$-0.29 \pm 0.03$\tnote{c} &
$0.17 \pm 0.10$\tnote{h} &
$-0.091 \pm 0.082$\tnote{l} &
$0.06 \pm 0.18$\tnote{n} &
$-0.39 \pm 0.03$\tnote{p} \\

$\log g$ (dex) &
$4.70 \pm 0.05$\tnote{c} &
$4.9 \pm 0.1$\tnote{h} &
$4.753 \pm 0.008$\tnote{l} &
$4.63 \pm 0.13$\tnote{n} &
$5.08 \pm 0.15$\tnote{q} \\

Luminosity ($L_\odot$) &
$0.099^{+0.010}_{-0.009}$\tnote{d} &
$0.0253 \pm 0.0021$\tnote{i} &
$0.0374 \pm 0.0021$\tnote{l} &
$\sim 0.029$\tnote{n} &
$0.00342 \pm 0.00003$\tnote{o} \\

Mass ($M_\odot$) &
$0.62 \pm 0.04$\tnote{c} &
$0.444 \pm 0.010$\tnote{j} &
$0.481 \pm 0.014$\tnote{l} &
$0.405 \pm 0.041$\tnote{n} &
$0.159 \pm 0.016$\tnote{o} \\

Radius ($R_\odot$) &
$0.58 \pm 0.03$\tnote{c} &
$0.450 \pm 0.013$\tnote{j} &
$0.478 \pm 0.021$\tnote{l} &
$0.453 \pm 0.014$\tnote{n} &
$0.1869 \pm 0.0012$\tnote{o} \\

$v\sin{i_\star}$ (km/s) &
$1.09^{+0.22}_{-0.20}$ \tnote{e} &
$< 2$ \tnote{k} &
$< 2$ \tnote{m} &
$<2$ \tnote{i} &
$< 2$ \tnote{q} \\

$v_{mac}$ (km/s) &
$2.162 \pm 2.632$\tnote{f} &
$2.625 \pm 2.320$\tnote{f} &
$2.432 \pm 2.355$\tnote{f} &
$2.438 \pm 2.240$\tnote{f} &
$3.763 \pm 1.887$\tnote{f} \\

$v_{mic}$ (km/s) &
$1.528 \pm 0.098$\tnote{f} &
$1.5 \pm 0.693$\tnote{f} &
$1.743 \pm 0.545$\tnote{f} &
$1.064 \pm 0.515$\tnote{f} &
$0.584 \pm 0.713$\tnote{f} \\

\bottomrule
\end{tabular*}

\begin{tablenotes}[para]\footnotesize
\item[a] \citet{Gaia_DR3} 
\item[b] \citet{Dressing_2019}
\item[c] \citet{Passegger2024} 
\item[d] \citet{PiauletGhorayeb_2024} 
\item[e] \citet{Friden_2026} 
\item[f] This work
\item[g] \citet{Benneke_2017} 
\item[h] \citet{Hejazi_2024} 
\item[i] \citet{Benneke_2019}
\item[j] \citet{Hardegree_2020}
\item[k] \citet{Sairam_2025} 
\item[l] \citet{Gore_2024} 
\item[m] \citet{Kossakowski_2021} 
\item[n] \citet{Melo2024}  
\item[o] \cite{Mann_2013}
\item[p] \citet{Jahandar2024}
\item[q] \citet{Maldonado2020}
\item[r] \citet{Reiners_2018}

\end{tablenotes}
\end{threeparttable}
\end{table*}

%% file: Tables/abundancestable.tex
\begin{table*}[htbp]
\centering
\caption{Abundances of Target Stars}
\label{tab:abundance_table}
\begin{tabular}{cccccc}
\hline
\hline
Element & GJ 9827 & K2-18 & TOI-1201 & TOI-1685 & Barnard's Star \\
\hline
$[\mathrm{C/H}]$  & $-0.203 \pm 0.121$ & $0.208 \pm 0.128$ & $0.020 \pm 0.108$ & $0.148 \pm 0.148$ & $-0.332 \pm 0.108$ \\
$[\mathrm{O/H}]$  & $-0.288 \pm 0.037$ & $0.134 \pm 0.092$ & $-0.119 \pm 0.082$ & $-0.001 \pm 0.138$ & $-0.308 \pm 0.075$ \\
$[\mathrm{Na/H}]$ & $-0.245 \pm 0.025$ & $0.248 \pm 0.036$ & $0.242 \pm 0.033$ & $0.230 \pm 0.053$ & $-0.449 \pm 0.082$ \\
$[\mathrm{Mg/H}]$ & $-0.223 \pm 0.179$ & $0.105 \pm 0.193$ & $-0.129 \pm 0.124$ & $-0.007 \pm 0.140$ & $-0.442 \pm 0.047$ \\
$[\mathrm{Al/H}]$ & $-0.292 \pm 0.027$ & $0.063 \pm 0.157$ & $-0.046 \pm 0.086$ & $0.071 \pm 0.143$ & $-0.288 \pm 0.089$ \\
$[\mathrm{Si/H}]$ & $-0.263 \pm 0.043$ & $0.217 \pm 0.183$ & $-0.106 \pm 0.152$ & $0.008 \pm 0.252$ & $-0.424 \pm 0.150$ \\
$[\mathrm{K/H}]$  & $-0.277 \pm 0.036$ & $0.093 \pm 0.110$ & $0.151 \pm 0.099$ & $0.191 \pm 0.144$ & $-0.570 \pm 0.100$ \\
$[\mathrm{Ca/H}]$ & $-0.435 \pm 0.029$ & $0.020 \pm 0.082$ & $-0.057 \pm 0.059$ & $0.007 \pm 0.085$ & $-0.474 \pm 0.144$ \\
$[\mathrm{Sc/H}]$ & $-0.287 \pm 0.066$ & -- & $-0.060 \pm 0.077$ & -- & -- \\
$[\mathrm{Ti/H}]$ & $-0.388 \pm 0.053$ & $0.029 \pm 0.121$ & $-0.178 \pm 0.105$ & $-0.017 \pm 0.163$ & $-0.673 \pm 0.065$ \\
$[\mathrm{V/H}]$  & -- & -- & $0.058 \pm 0.038$ & $0.060 \pm 0.109$ & -- \\
$[\mathrm{Cr/H}]$ & $-0.407 \pm 0.046$ & $0.151 \pm 0.084$ & $-0.031 \pm 0.070$ & $0.009 \pm 0.116$ & $-0.353 \pm 0.093$ \\
$[\mathrm{Mn/H}]$ & $-0.254 \pm 0.052$ & $0.401 \pm 0.126$ & $0.088 \pm 0.106$ & $0.343 \pm 0.150$ & $-0.676 \pm 0.085$ \\
$[\mathrm{Fe/H}]$ & $-0.360 \pm 0.037$ & $-0.048 \pm 0.098$ & $-0.144 \pm 0.091$ & $-0.078 \pm 0.138$ & $-0.722 \pm 0.085$ \\
$[\mathrm{Ni/H}]$ & $-0.325 \pm 0.082$ & -- & $-0.039 \pm 0.065$ & -- & -- \\
\hline
$[\mathrm{M/H}]$\footnote{Calculated using the formalism from \cite{Hinkel2022}} & $-0.26 \pm 0.05$ & $0.16 \pm 0.07$ & $-0.06 \pm 0.06$ & $ 0.06 \pm 0.09$ & $-0.33 \pm 0.06$ \\
Avg. $[\mathrm{M/H}]$\footnote{Calculated as an average} & $-0.29 \pm 0.07$ & $0.14 \pm 0.12$ & $-0.02 \pm 0.11$ & $0.07 \pm 0.12$ & $-0.48 \pm 0.15$\\
Lit. $[\mathrm{M/H}]$\footnote{See Table \ref{tab:stellarparams} for citations} & $-0.29 \pm 0.03$ & $0.17 \pm 0.10$ & $-0.12 \pm 0.08$ & $0.06 \pm 0.18$ & $-0.48 \pm 0.04$\\
$[\alpha/\mathrm{Fe}]$ & $0.08 \pm 0.05$ & $0.19 \pm 0.13$ & $0.03 \pm 0.12$ & $0.09 \pm 0.18$ & $0.40 \pm 0.11$ \\ 
Avg. $[\alpha/\mathrm{Fe}]$ & $0.04 \pm 0.05$ & $0.15 \pm 0.12$ & $0.03 \pm 0.10$ & $0.08 \pm 0.16$ & $0.29 \pm 0.10$ \\ 
\hline

\end{tabular}
\end{table*}

%% file: Tables/abundanceratios.tex
\begin{table*}[htbp]
\centering
\caption{Elemental Abundance Ratios}
\label{tab:ratio_table}
\footnotesize
\begin{tabular}{cccccccccc}
\hline
\hline
Element & GJ 9827 & K2-18 & TOI-1201 & TOI-1685 & Barnard's Star & Sun\footnote{Photospheric abundance ratios from \citet{Asplund2009}.} & J25 M dwarfs \footnote{\citet{Jahandar_2025}} & Brewer 2016 (FGK)\footnote{\citet{BrewerFischerCoolStars2016}}\\
\hline 
C/O & $0.67^{+0.22}_{-0.17}$ & $0.65^{+0.28}_{-0.20} $ & $0.76^{+0.28}_{-0.21} $ & $0.78^{+0.46}_{-0.29}$ & $0.52^{+0.18}_{-0.14} $ & 0.55 $\pm$ 0.17 & 0.82 $\pm$ 0.21 & 0.45 $\pm$ 0.10\\
Fe/Mg & $0.58^{+0.30}_{-0.20}$ & $0.56^{+0.36}_{-0.22} $ & $0.77^{+0.32}_{-0.23} $ & $0.68^{+0.28}_{-0.24}$ & $0.42^{+0.10}_{-0.08} $ & 0.79 $\pm$ 0.14 & 0.69 $\pm$ 0.26 & 0.86 $\pm$ 0.13\\
Mg/Si & $1.35^{+0.70}_{-0.47}$ & $0.95^{+0.80}_{-0.43} $ & $1.16^{+0.66}_{-0.42} $ & $1.19^{+1.10}_{-0.57}$ & $1.18^{+0.51}_{-0.36} $ & 1.23 $\pm$ 0.12 & 1.62 $\pm$ 0.74 & 1.25 $\pm$ 0.18\\
Fe/O & $0.05^{+0.01}_{-0.01}$ & $0.04^{+0.02}_{-0.01} $ & $0.06 \pm 0.02 $ & $0.05^{+0.03}_{-0.02}$ & $0.03 \pm 0.01 $ & 0.07 $\pm$ 0.02 & 0.09 $\pm$ 0.03  & 0.06 $\pm$ 0.02
\end{tabular}
\end{table*}

%% file: Tables/abundanceerrors.tex
\begin{table}[htbp]
\centering
\caption{Complete Record of Abundances and Errors}
\label{tab:Error_table}
\begin{tabular}{cccccccccc}
\hline
Element & \# of lines & [X/H] & $\sigma_{\rm rand}$ & $\sigma_{\rm Teff}$ & $\sigma_{\rm [M/H]}$ & $\sigma_{\log{g}}$ & $\sigma_{\rm(v_{mac})}$ & $\sigma_{\rm(v_{mic})}$ & $\sigma_{\rm total}$ \\
&& (dex) & (dex)& (dex)& (dex)& (dex)& (dex)& (dex)& (dex)  \\
\hline
\multicolumn{10}{c}{{\it GJ 9827}} \\
\hline
C (CO) & 176 & -0.203 & 0.004 & 0.002 & 0.003 & 0.030 & 0.116 & 0.012 & 0.121 \\
O (OH) & 35 & -0.288 & 0.002 & 0.004 & 0.025 & 0.004 & 0.027 & 0.001 & 0.037 \\
Na & 6 & -0.245 & 0.023 & 0.001 & 0.001 & 0.006 & 0.009 & 0.002 & 0.025 \\
Mg & 7 & -0.223 & 0.057 & 0.025 & 0.084 & 0.066 & 0.105 & 0.078 & 0.179 \\
Al & 11 & -0.292 & 0.024 & 0.004 & 0.004 & 0.007 & 0.008 & 0.002 & 0.027 \\
Si & 32 & -0.263 & 0.020 & 0.017 & 0.011 & 0.021 & 0.025 & 0.002 & 0.043 \\
K & 4 & -0.277 & 0.032 & 0.002 & 0.004 & 0.007 & 0.013 & 0.005 & 0.036 \\
Ca & 17 & -0.435 & 0.016 & 0.001 & 0.001 & 0.004 & 0.023 & 0.002 & 0.029 \\
Ti & 57 & -0.388 & 0.014 & 0.004 & 0.005 & 0.012 & 0.048 & 0.011 & 0.053 \\
Sc & 2 & -0.287 & 0.013 & 0.005 & 0.006 & 0.018 & 0.061 & 0.002 & 0.066 \\
Cr & 28 & -0.407 & 0.014 & 0.003 & 0.004 & 0.015 & 0.040 & 0.004 & 0.046 \\
Mn & 9 & -0.254 & 0.043 & 0.004 & 0.011 & 0.009 & 0.022 & 0.012 & 0.052 \\
Fe & 80 & -0.360 & 0.017 & 0.008 & 0.006 & 0.020 & 0.023 & 0.002 & 0.037 \\
Ni & 6 & -0.325 & 0.062 & 0.008 & 0.005 & 0.024 & 0.048 & 0.000 & 0.082 \\
\hline
\multicolumn{10}{c}{{\it K2-18}} \\
\hline
C (CO) & 94 & 0.208 & 0.016 & 0.009 & 0.056 & 0.029 & 0.077 & 0.079 & 0.128 \\
O (OH) & 47 & 0.134 & 0.005 & 0.008 & 0.086 & 0.007 & 0.017 & 0.025 & 0.092 \\
Na & 4 & 0.248 & 0.013 & 0.024 & 0.004 & 0.015 & 0.017 & 0.001 & 0.036 \\
Mg & 7 & 0.105 & 0.077 & 0.150 & 0.011 & 0.069 & 0.062 & 0.014 & 0.193 \\
Al & 5 & 0.063 & 0.051 & 0.117 & 0.053 & 0.064 & 0.038 & 0.012 & 0.157 \\
Si & 4 & 0.217 & 0.034 & 0.167 & 0.051 & 0.037 & 0.012 & 0.020 & 0.183 \\
K & 5 & 0.093 & 0.032 & 0.081 & 0.039 & 0.022 & 0.007 & 0.051 & 0.110 \\
Ca & 14 & 0.020 & 0.059 & 0.036 & 0.029 & 0.012 & 0.026 & 0.019 & 0.082 \\
Ti & 37 & 0.029 & 0.021 & 0.028 & 0.066 & 0.023 & 0.042 & 0.082 & 0.121 \\
Cr & 18 & 0.151 & 0.017 & 0.041 & 0.038 & 0.025 & 0.043 & 0.033 & 0.084 \\
Mn & 4 & 0.401 & 0.048 & 0.027 & 0.068 & 0.006 & 0.006 & 0.090 & 0.126 \\
Fe & 28 & -0.048 & 0.027 & 0.047 & 0.040 & 0.024 & 0.032 & 0.059 & 0.098 \\
\hline
\end{tabular}
\end{table}

%% file: Tables/abundanceerrors_cont.tex
\addtocounter{table}{-1}
\begin{table}[htbp]
\centering
\caption{Complete Record of Abundances and Errors (continued)}
\label{tab:Fe_lines}
\begin{tabular}{cccccccccc}
\hline
Element & \# of lines & [X/H] & $\sigma_{\rm rand}$ & $\sigma_{\rm Teff}$ & $\sigma_{\rm [M/H]}$ & $\sigma_{\log{g}}$ & $\sigma_{\rm(v_{mac})}$ & $\sigma_{\rm(v_{mic})}$ & $\sigma_{\rm Total}$ \\
&& (dex) & (dex)& (dex)& (dex)& (dex)& (dex)& (dex)& (dex)  \\
\multicolumn{10}{c}{{\it TOI-1201}} \\
\hline
C (CO) & 116 & 0.020 & 0.011 & 0.003 & 0.050 & 0.002 & 0.063 & 0.072 & 0.108 \\
O (OH) & 46 & -0.119 & 0.005 & 0.003 & 0.063 & 0.003 & 0.030 & 0.043 & 0.082 \\
Na & 3 & 0.242 & 0.014 & 0.023 & 0.003 & 0.001 & 0.018 & 0.002 & 0.033 \\
Mg & 9 & -0.129 & 0.040 & 0.096 & 0.031 & 0.007 & 0.035 & 0.049 & 0.124 \\
Al & 6 & -0.046 & 0.023 & 0.071 & 0.030 & 0.004 & 0.026 & 0.016 & 0.086 \\
Si & 8 & -0.106 & 0.045 & 0.127 & 0.065 & 0.004 & 0.023 & 0.008 & 0.152 \\
K & 6 & 0.151 & 0.058 & 0.052 & 0.021 & 0.001 & 0.016 & 0.055 & 0.099 \\
Ca & 18 & -0.057 & 0.035 & 0.018 & 0.015 & 0.001 & 0.015 & 0.039 & 0.059 \\
Ti & 45 & -0.178 & 0.021 & 0.010 & 0.052 & 0.002 & 0.039 & 0.079 & 0.105 \\
Sc & 2 & -0.060 & 0.061 & 0.004 & 0.036 & 0.003 & 0.008 & 0.028 & 0.077 \\
Cr & 17 & -0.031 & 0.018 & 0.035 & 0.037 & 0.003 & 0.037 & 0.026 & 0.070 \\
Mn & 6 & 0.088 & 0.051 & 0.033 & 0.053 & 0.002 & 0.031 & 0.062 & 0.106 \\
V & 1 & 0.058 & 0.000 & 0.021 & 0.027 & 0.003 & 0.017 & 0.001 & 0.038 \\
Fe & 37 & -0.144 & 0.026 & 0.054 & 0.045 & 0.003 & 0.020 & 0.047 & 0.091 \\
Ni & 1 & -0.039 & 0.000 & 0.043 & 0.037 & 0.004 & 0.032 & 0.003 & 0.065 \\
\hline
\multicolumn{10}{c}{{\it TOI-1685}} \\
\hline
C (CO) & 51 & 0.148 & 0.015 & 0.010 & 0.101 & 0.029 & 0.081 & 0.063 & 0.148 \\
O (OH) & 24 & -0.001 & 0.005 & 0.025 & 0.115 & 0.040 & 0.039 & 0.044 & 0.138 \\
Na & 3 & 0.230 & 0.024 & 0.036 & 0.015 & 0.002 & 0.025 & 0.008 & 0.053 \\
Mg & 5 & -0.007 & 0.075 & 0.102 & 0.027 & 0.039 & 0.011 & 0.034 & 0.140 \\
Al & 3 & 0.071 & 0.034 & 0.117 & 0.068 & 0.014 & 0.004 & 0.025 & 0.143 \\
Si & 3 & 0.008 & 0.025 & 0.208 & 0.126 & 0.055 & 0.019 & 0.014 & 0.252 \\
K & 3 & 0.191 & 0.103 & 0.069 & 0.039 & 0.044 & 0.011 & 0.042 & 0.144 \\
Ca & 8 & 0.007 & 0.033 & 0.028 & 0.055 & 0.017 & 0.029 & 0.034 & 0.085 \\
Ti & 33 & -0.017 & 0.016 & 0.029 & 0.132 & 0.018 & 0.033 & 0.082 & 0.163 \\
Cr & 16 & 0.009 & 0.018 & 0.055 & 0.076 & 0.031 & 0.048 & 0.032 & 0.116 \\
Mn & 4 & 0.343 & 0.033 & 0.044 & 0.118 & 0.002 & 0.025 & 0.071 & 0.150 \\
V & 1 & 0.060 & 0.000 & 0.051 & 0.072 & 0.047 & 0.044 & 0.002 & 0.109 \\
Fe & 14 & -0.078 & 0.018 & 0.058 & 0.098 & 0.022 & 0.035 & 0.063 & 0.138 \\
\hline
\multicolumn{10}{c}{{\it Barnard's Star}} \\
\hline
C (CO) & 59 & -0.332 & 0.021 & 0.003 & 0.038 & 0.022 & 0.080 & 0.053 & 0.108 \\
O (OH) & 27 & -0.308 & 0.010 & 0.016 & 0.011 & 0.062 & 0.025 & 0.027 & 0.075 \\
Na & 2 & -0.449 & 0.041 & 0.009 & 0.017 & 0.067 & 0.009 & 0.003 & 0.081 \\
Mg & 1 & -0.442 & 0.000 & 0.012 & 0.036 & 0.017 & 0.021 & 0.005 & 0.047 \\
Al & 5 & -0.288 & 0.051 & 0.010 & 0.023 & 0.054 & 0.041 & 0.011 & 0.089 \\
Si & 1 & -0.424 & 0.000 & 0.018 & 0.052 & 0.136 & 0.030 & 0.004 & 0.150 \\
K & 3 & -0.570 & 0.008 & 0.047 & 0.009 & 0.085 & 0.010 & 0.014 & 0.100 \\
Ca & 9 & -0.474 & 0.131 & 0.009 & 0.029 & 0.041 & 0.031 & 0.002 & 0.144 \\
Ti & 26 & -0.673 & 0.023 & 0.015 & 0.038 & 0.023 & 0.029 & 0.026 & 0.065 \\
Cr & 10 & -0.353 & 0.053 & 0.014 & 0.020 & 0.068 & 0.025 & 0.002 & 0.093 \\
Mn & 2 & -0.676 & 0.076 & 0.000 & 0.025 & 0.011 & 0.022 & 0.013 & 0.085 \\
Fe & 15 & -0.722 & 0.062 & 0.006 & 0.035 & 0.027 & 0.031 & 0.022 & 0.085 \\
\hline
\end{tabular}
\end{table}

%% file: Tables/linelist_shortened.tex
\begin{table}[htbp]
\centering
\caption{Complete linelist and line-by-line abundances}
\label{tab:complete_lines}
\begin{tabular}{ccccccc}
\hline
Wavelength [nm] & Species & GJ 9827 & K2-18 & TOI-1201 & TOI-1685 & Barnard's Star \\
\hline
1078.19722 & Al 1 &  &  & -0.02 &  &  \\
1087.29300 & Al 1 & -0.11 &  &  &  &  \\
1089.16740 & Al 1 & -0.15 &  &  &  &  \\
1125.31943 & Al 1 & -0.33 &  &  &  & -0.41 \\
1125.48831 & Al 1 & -0.26 &  & -0.03 &  & -0.52 \\
1312.34148 & Al 1 & -0.36 & 0.10 & -0.13 & 0.06 & -0.25 \\
1315.07514 & Al 1 & -0.39 & 0.05 & -0.03 & 0.14 & -0.29 \\
1671.89078 & Al 1 & -0.31 & 0.17 & -0.08 & -0.00 &  \\
1675.05117 & Al 1 & -0.39 & -0.19 &  &  & -0.15 \\
1676.33154 & Al 1 & -0.26 & 0.15 & 0.05 &  &  \\
2109.29737 & Al 1 & -0.19 &  &  &  &  \\
2116.37094 & Al 1 & -0.24 &  &  &  &  \\
2292.92232 & CO 1 & -0.19 &  & 0.22 & 0.16 & -0.40 \\
2293.01808 & CO 1 & -0.20 & 0.22 &  & 0.16 & -0.39 \\
2293.14330 & CO 1 & -0.18 &  &  &  & -0.50 \\
2293.31778 & CO 1 & -0.19 & 0.22 &  & 0.11 & -0.35 \\
2293.48421 & CO 1 &  & 0.30 &  &  &  \\
2293.53404 & CO 1 & -0.23 & 0.14 &  &  & -0.44 \\
2293.72435 & CO 1 & -0.20 & 0.09 & 0.07 & -0.05 &  \\
2293.78470 & CO 1 & -0.26 & 0.15 & -0.10 & 0.17 & -0.31 \\
2294.00382 & CO 1 & -0.25 & 0.31 & -0.02 & 0.06 & -0.27 \\
2294.07627 & CO 1 &  & 0.28 & 0.06 &  & -0.16 \\
2294.32729 & CO 1 & -0.18 &  & -0.11 &  & -0.34 \\
2294.40766 & CO 1 & -0.24 & 0.27 & -0.00 & 0.20 &  \\
2294.69197 & CO 1 & -0.30 & 0.07 & 0.02 &  & -0.30 \\
2294.77774 & CO 1 & -0.21 & 0.33 & -0.09 &  & -0.31 \\
... & ... & ... & ... & ... & ... & ... \\
\hline
\end{tabular}
\end{table}